%% file: main.tex
\documentclass[twocolumn, reprint, letterpaper,aps,prl,superscriptaddress,nofootinbib,floatfix]{revtex4-2}

\usepackage{graphicx}
\usepackage{dcolumn}
\usepackage{bm}
\usepackage{hyperref}
\usepackage{amsmath}
\usepackage{tikz}
\usepackage{orcidlink}

\def\MJD{\textsc{Majorana Demonstrator}}
\def\dem{\textsc{Demonstrator}}

\def\bb{$\beta\beta$}
\def\znbb{$0\nu\beta\beta$}
\def\tnbb{$2\nu\beta\beta$}

\def\es{E.S.}
\def\gs{G.S.}

\graphicspath{{./figs/}}

\begin{document}

\title{Final Results of the \bfseries{M{\small AJORANA} D{\small EMONSTRATOR}}'s Search for Double-Beta Decay of $^{76}$Ge to Excited States of $^{76}$Se}
\include{author_list.tex}
\date{\today}
\begin{abstract}
$^{76}$Ge can $\beta\beta$\ decay into three possible excited states of $^{76}$Se, with the emission of two or, if the neutrino is Majorana, zero neutrinos.
None of these six transitions have yet been observed.
The \MJD\ was designed to study $\beta\beta$\ decay of $^{76}$Ge using a low background array of high purity germanium detectors.
With 98.2~kg-y of isotopic exposure, the \dem\ sets the strongest half-life limits to date for all six transition modes.
For \tnbb\ to the $0^+_1$ state of $^{76}$Se, this search has begun to probe for the first time half-life values predicted using modern many-body nuclear theory techniques, setting a limit of $T_{1/2}>1.5\times10^{24}$~y (90\% CL).
\end{abstract}

\footnote{Notice: This manuscript has been authored by UT-Battelle, LLC, under contract DE-AC05-00OR22725 with the US Department of Energy (DOE). The US government retains and the publisher, by accepting the article for publication, acknowledges that the US government retains a nonexclusive, paid-up, irrevocable, worldwide license to publish or reproduce the published form of this manuscript, or allow others to do so, for US government purposes. DOE will provide public access to these results of federally sponsored research in accordance with the DOE Public Access Plan (\url{https://www.energy.gov/doe-public-access-plan}).}

\maketitle

Double-beta (\bb) decay is a rare second-order weak nuclear process in which two neutrons simultaneously decay to two protons and emit two electrons.
\bb\ decay was predicted by Goeppert-Mayer to occur in even-even nuclei in which a single $\beta$\ decay is forbidden~\cite{GoeppertMayer1935}.
Furthermore, if the neutrino is a Majorana fermion, meaning it is its own antiparticle~\cite{Majorana1937}, then neutrinoless double-beta decay (\znbb) can occur~\cite{Furry1939}.
\bb\ decay with the emission of two neutrinos (\tnbb) has been measured in 11~isotopes, with half-lives in a range of $10^{18}-10^{24}$~yr~\cite{Barabash2020}.
\znbb\ has not been observed, but its discovery would prove that the neutrino is a Majorana fermion~\cite{Schechter1982}, provide an example of lepton number violation in nature, and might provide a mechanism for the generation of the observed matter-antimatter asymmetry in the universe~\cite{Sakharov1967, Fukugita1986}.
A robust experimental program is searching for \znbb\ in a variety of isotopes~\cite{Dolinski2019, Gomez-Cadenas2023, Barabash2023, Agostini2023}.

\bb\ decay can cause a transition of parent nuclei to daughters in either the ground state (\gs) or an energetically allowed excited state (\es)~\cite{Belli2020}.
Decays to \es s have a lower Q-value than decays to the \gs\ and promptly emit one or more $\gamma$\ rays.
The branching ratios are suppressed for \es\ decays relative to \gs\ decays, due to the smaller phase space of the decay.
To date, only $\beta\beta$ transitions to the first $0^+$ \es\ of two isotopes have been observed, in $^{100}$Mo ($T_{1/2}=6.7^{+0.5}_{-0.4}\times10^{20}$~y)~\cite{Barabash1995,  Barabash1999, DeBraeckeleer2001, Arnold2007, Kidd2009, Belli2014, Arnold2014, Augier2023} and $^{150}$Nd ($T_{1/2}=1.18^{+0.23}_{-0.20}\times10^{20}$~y)~\cite{Barabash2004, Barabash2009, Kidd2014, Polischuk2021, Aguerre2023}.

Applying Fermi's golden rule and the closure approximation, we express the half-life for \tnbb\ as:
\begin{equation} \label{eq:hl_tnbb}
    T_{1/2}^{-1} = G^{2\nu}_{s}\cdot(g_A^{eff,2\nu})^4\cdot|M^{2\nu}_{s}|^2
\end{equation}
where $G^{2\nu}_{s}$ is the phase space factor (PSF), which depends on the daughter nuclear state $s$, $(g_A^{eff,2\nu})^2$ is the axial vector coupling constant with an empirical quenching term applied, and $|M^{2\nu}_{s}|$ is the nuclear matrix element (NME).
The PSF can be accurately calculated~\cite{Kotila2012, Neacsu2016, Stoica2019}, but there is a large uncertainty on $(g_A^{eff,2\nu})^4\cdot|M^{2\nu}_{s}|^2$~\cite{Suhonen2017}; this means that half-life measurements of \tnbb\ to the \gs\ and \es s\ serve as useful tests of nuclear many-body models used to compute the NME.
In addition, the NME for \tnbb\ transitions to $2^+$ states is sensitive to a bosonic component of the neutrino wave function~\cite{Dolgov2005, Barabash2007}.

For \znbb\ dominated by light neutrino exchange, the half-life is expressed as:
\begin{equation}
\label{eq:hl_znbb}
    T_{1/2}^{-1}=G^{0\nu}_{s} \cdot (g^{eff, 0\nu}_{A})^4|M^{0\nu}|^2\langle m_{\beta\beta}\rangle^2
\end{equation}
where $m_{\beta\beta}$ is the effective Majorana mass of the electron neutrino.
Given an accurate calculation of $(g^{eff, 0\nu}_{A})^4|M^{0\nu}|^2$, a measurement of the \znbb\ half-life would provide information about the neutrino mass and Majorana CP-phases~\cite{Engel2017}.
Furthermore, the branching ratio for \bb\ decay modes to \es s can vary depending on the physics mechanisms; this means that a measurement of \znbb\ to an \es\ of the daughter nucleus could help inform how to extend the Standard Model to accommodate Majorana neutrinos~\cite{Simkovic2002}.

$^{76}$Ge is a promising isotope with an active experimental program for measuring \bb\ decay~\cite{Legend:pCDR}.
Arrays of high purity germanium (HPGe) detectors that are isotopically enriched in $^{76}$Ge can achieve high detection efficiency, an ultra-low background rate, and excellent energy resolution. 
The \MJD~\cite{mjd:instrumentation2025}, which operated HPGe detectors in vacuum, and GERDA~\cite{gerda2018}, which operated HPGe detectors submerged in liquid argon instrumented to act as an active veto, both recently completed their data-taking campaigns and achieved the two lowest background indices and best energy resolutions in their searches for \znbb\ among experiments performed to date~\cite{mjd2023, gerda2020}.

$^{76}$Ge can decay into three \es s of $^{76}$Se with a decay structure shown in Fig.~\ref{fig:level_diagram}; none of these transitions have been observed before.
Searches for \es\ decay modes were performed by both the \MJD~\cite{mjd:ES2021} and GERDA~\cite{gerda:ES} by searching for peaks produced when deexcitation $\gamma$ rays escape the detector of origin and are fully absorbed in a second HPGe detector.

\begin{figure}
    \input{level_diagram.tex}
    \caption{\label{fig:level_diagram}
        Level diagram for $\beta\beta$\ decay of $^{76}$Ge to $^{76}$Se.
    }
\end{figure}

\bigskip
The \MJD\ searched for \znbb\ and \bb\ decay to \es s using an array of HPGe detectors.
The experiment deployed two modules, each consisting of 29~HPGe detectors operated in a separate vacuum cryostat.
The modules were constructed from ultra-low background materials~\cite{mjd:efcu, mjd:assay} and placed in a low-background passive shield, surrounded by a muon veto with nearly $4\pi$-coverage~\cite{mjd:muonveto, mjd:muonflux}.
For each module, a $^{228}$Th line source stored outside of this shield was deployed once per week along a helical track surrounding the cryostat to calibrate the detectors~\cite{mjd:calibration, mjd:calibrationprocedure}.
Signals from the HPGe detectors were digitized~\cite{mjd:electronics}, with each channel triggered independently with typical energy thresholds of $<1$~keV and a dynamic range up to 10~MeV.
Waveforms were stored on disk and reanalyzed to calculate standard data cleaning cuts that reject non-physical events and keep $>99.9\%$ of physical events~\cite{mjd2019}, PSD parameters~\cite{mjd:avse, mjd:dcr}, and more accurate energies with corrections for digitizer non-linearity~\cite{mjd:adcnonlin} and charge-trapping~\cite{mjd:chargetrapping}.
The experiment was located at the 4850~ft level (4300 m.w.e.) in the Davis campus of the Sanford Underground Research Facility, in Lead, SD~\cite{surf}.

The \dem\ utilized p-type HPGe detector geometries with a $p^+$-type point-like electrode on one face, and an $n^+$-type electrode on the other surfaces.
Three geometries were deployed, including p-type point-contact (PPC) detectors~\cite{Barbeau2007} and inverted coaxial point contact (ICPC) detectors~\cite{Cooper2011}, each enriched to 87--88\% in $^{76}$Ge, and BEGe\texttrademark\ detectors~\cite{Canberra:bege} with a natural isotopic abundance of 7.8\% $^{76}$Ge.
Module~1 began operation in its low background configuration in December 2015, and Module~2 began in August 2016.
For this analysis, we divide data into five datasets, listed in Tab.~\ref{tab:details}, based on which sets of detectors were deployed; these are combinations of the 13~datasets described in Ref.~\cite{mjd2023}.
We exclude a period from Oct.~2016 to Jan.~2017 with higher electronic noise due to sub-optimal grounding.
During most of its operation, blinding was applied in alternating cycles of 31~h of open data followed by 93~h of blind data~\cite{mjd2019}; this enables us to optimize our analysis while keeping sensitivity using only open data below previously published results.

The detection signature used to identify \bb\ to \es s is energy peaks created by the full absorption of a gamma in a detector different from the site of the decay.
This signature directly takes advantage of the \dem's strength in peak searches, which derives from its excellent energy resolution.
A typical event following this signature involves multiple detector hits in coincidence, so we accept events with a detector multiplity $\geq2$.
\bb\ to \es\ events with multiplicity~1 will produce a broad spectral signal from the sum of the gamma ray and beta decay energies, with a large background from the spectral signal from \tnbb\ to the \gs\ resulting in low sensitivity from this signal compared to our selected signature; in addition, analysis of broad spectral features is subject to systematic uncertainties from the modelling of backgrounds to which this peak search analysis is immune.

Detector hits that fall within a rolling $4~\mu$s window are combined into events, with multiplicity defined as the number of HPGe detector hits in a single event.
An event including any hit that fails the data cleaning cuts is rejected.
In addition, we reject events within 20~ms before and 1~s after a muon event, removing $<$1\% of events~\cite{mjd:muonflux, mjd:muonactivity}.
High multiplicity events also contain valuable information from coincident detectors that we use to design multiple background rejection cuts.
These cuts were developed and optimized using open data and simulations.
This analysis technique was also used in Ref.~\cite{mjd:ES2021}; since then, refinements have been made to further improve sensitivity, which will be noted.
This text focuses on \bb\ to the $0^+_1$ \es, the likeliest transition to be observed; we describe any differences with other \es\ transitions.

We use simulations to estimate the detection efficiency of these peaks including the effect of background cuts, and to optimize the tradeoff between signal sacrifice and background reduction, boosting our sensitivity.
MaGe~\cite{mjd:mage} is a Geant4~\cite{geant2003} based software library that implements the full as-built geometry for each experimental configuration of the \MJD\ and produces Monte-Carlo simulations of a variety of physical processes.
To generate \bb\ decays to \es s, the DECAY0~\cite{Ponkratenko2000} library was used, with modifications described in Ref.~\cite{mjd:ES2021}.
Other radioactive backgrounds and calibration source events were generated with the standard radioactive decay module built into Geant4.

Geant4 step data is post-processed to simulate the observables produced by HPGe detectors.
This stage simulates the effect of dead time from detectors that are disabled or unstable, from data cleaning cuts, and from hardware retriggering by randomly rejecting detector hits in proportion with the time spent in that configuration.
We simulate incomplete charge collection in transition dead layers within $\sim$1~mm of the $n^+$ detector surfaces by reducing the energy deposited by steps in these regions.
One hundred separate sets of post-processed simulations are produced for each dataset and \es\ decay mode, varying the transition dead layer parameters to study associated systematic effects.
In addition, a background model simulation is produced by sampling from about 100 post-processed simulations of a variety of isotopes in different hardware components, in proportion with the fitted activities from Ref.~\cite{reine:2023}.

We apply a sequence of background reduction cuts, determined based on our simulations to improve the sensitivity of the experiment.
The ``Enriched Source Detector Cut" rejects hits that are not in coincidence with an enriched detector.
The ``Hot Detector Cut" rejects events that include one of two detectors closest to the Module~1 crossarm; this cut was not included in Ref.~\cite{mjd:ES2021}
These detectors have significantly elevated background rates consistent with $^{232}$Th progeny in a cavity in the interface between the Module~1 cryostat cold plate and crossarm~\cite{haufe:2023, reine:2023}.
For \bb\ to the $2^+_1$ \es\ and $2^+_2$ \es\ with the emision of a 1216~keV $\gamma$\ ray, we additionally require that an event has a multiplicity of exactly 2, since these modes only emit a single $\gamma$\ ray.

Because most $\gamma$\ rays that are fully absorbed inside a detector Compton scatter at least once, we select for multi-site hits in the gamma peak with the $AvsE$ PSD parameter~\cite{mjd:avse}.
$AvsE$ compares the current amplitude ($A$) and energy ($E$) of a pulse to tag hits as either single-site or multi-site; multi-site hits usually have a lower $A$ for a given $E$ in point-contact HPGe detectors relative to single-site hits.
$AvsE$ is corrected for correlations with drift time and energy, and calibrated to 90\% of single-site hits in the $^{208}$Tl double-escape peak (DEP)~\cite{mjd2023}.
We measure the multi-site acceptance for full energy peaks (FEPs) in 16~peaks from $^{228}$Th calibration data between 400~keV and 1700~keV, and model dependence on energy ($E_{FEP}$) as
\begin{equation}
    \varepsilon_{AvsE}(E_{FEP}, p_0, p_1) = p_0 - \frac{p_1}{E_{FEP}}
\end{equation}
For each dataset, we optimize the parameters $p_0\simeq 0.92$ and $p_1\simeq 3.4\times10^4$~keV combining all detectors and calibration runs, and the fit performs well with a $\chi^2$ value ranging from 10-20 for 14 d.o.f.
We check the acceptance of the 511~keV annihilation peak in coincidence with a DEP or single-escape peak (SEP) events for differences based on whether a $\gamma$ originates inside detectors instead of calibration track; this is consistent with the above model.
We also check for variance over time in the DEP, SEP, and Compton continuum acceptances, and for variance among the acceptances across many detectors of the 583~keV FEP; these are used to calculate systematic uncertainty terms, with the dominant uncertainty arising from variance over time.
Based on this, we measure an acceptance of $80.9\pm0.2\%$ for the 559~and 563~keV $\gamma$\ rays from the $0^+_1$ \es\ of $^{76}$Se.
This cut was not used in Ref.~\cite{mjd:ES2021}.

Lastly, we apply the Sum- and Coincident-Energy cuts, which reject events where either the sum over all hits or any hit in a coincident detector fall within a set of energy ranges.
These cuts target multi-detector events from $\gamma$\ ray cascades and Compton-scattered $\gamma$ rays from common backgrounds, respectively.
The energy ranges were determined algorithmically to optimize the discovery sensitivity of the experiment, based on the signal and background acceptance efficiencies determined using simulations of the \es\ modes and of the background model.
The events were binned both by sum- and coincident hit energies, and bins were added to the cut if doing so improved the sensitivity.
To avoid statistical biases towards cutting statistical fluctuations in the simulations, a new energy range was only introduced to the cut if we estimated it to have a $>$97\% chance of improving sensitivity.
For the \znbb\ decay to the $2^+_1$ and the single-$\gamma$ branch of the $2^+_2$ mode, we expect one event at the \es 's $Q_{\beta\beta}$ and one at the $\gamma$-ray energy; instead of this algorithm  we simply apply a coincident energy cut around the $Q_{\beta\beta}$-value.

\begin{figure}
    \centering
    \includegraphics[width=0.48\textwidth]{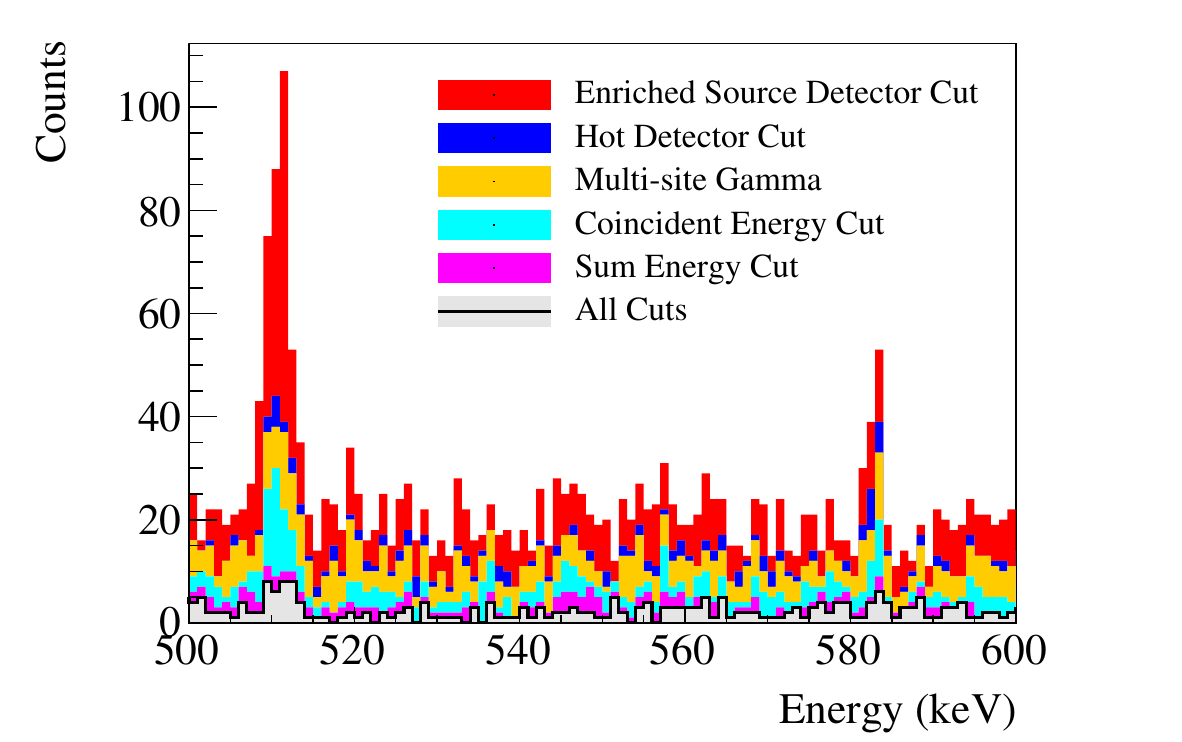}
    \caption{\label{fig:datacuts}
        Energy spectrum for high multiplicity events from full dataset, with background reduction cuts for \tnbb\ to the $0^+_1$ \es\ of $^{76}$Se applied in sequence.
    }
\end{figure}
\begin{table}[]
    \centering
    \begin{tabular}{l|c|c}
        Cut Description & $\varepsilon_{signal}$ & $\varepsilon_{background}$ \\
        \hline
        Gamma FEP Efficiency & 5.4\% & -- \\
        Multiplicity $\geq2$ & 71.5\% & 8.0\% \\
        Enriched Source Detector Cut & 97.5\% & 63.6\% \\
        Hot Detector Cut & 97.4\% & 88.9\% \\
        Multi-site Gamma & 80.9\% & 56.2\% \\
        Coincident Energy Cut & 82.6\% & 54.8\% \\
        Sum Energy Cut & 88.8\% & 56.9\% \\
        \hline
        Total & 2.2\% & 0.8\%
    \end{tabular}
    \caption{Signal acceptance $\varepsilon_{signal}$ and background acceptance $\varepsilon_{background}$ for \tnbb\ to the $0^+_1$ \es\ of $^{76}$Se, evaluated on cuts applied in the listed sequence. $\varepsilon_{signal}$ is averaged over datasets, and $\varepsilon_{background}$ is calculated from the events rejected in the fit window for all datasets (see Fig.~\ref{fig:datacuts}).}
    \label{tab:cutsummary}
\end{table}

The effect of the cuts for \bb\ to the $0^+_1$ \es\ is shown in Tabs.~\ref{tab:cutsummary}/\ref{tab:details} and Fig.~\ref{fig:datacuts}, and the final detection efficiency for each decay mode can be seen in Tab.~\ref{tab:results}.
These were determined by measuring the efficiency in simulations and multiplying the $AvsE$ efficiency determined for each energy peak.
Systematic uncertainty due to dead layer thickness and dead time is derived from the variance introduced by changing the simulation post-processing parameters, measured to be $0.06\%$.
In addition, uncertainty in the spectral shape from DECAY0 was estimated at $0.01\%$ by performing a Kolmogorov-Smirnov test comparing the \gs\ spectrum to a more precise determination from Ref.~\cite{Kotila2012}.
Finally, we validate the simulations using DEPs in coincidence with full absorption of a 511~keV annihilation $\gamma$ as proxies for \bb\ to \es s, since these $\gamma$s originate inside detectors.
To do this, we use data collected using a $^{56}$Co line source that was inserted into each calibration track for one week; this source emits many high-energy $\gamma$ rays, and we used 6 DEPs and 7 SEPs.
We compared the measured and simulated ratios of the peak amplitudes for hits in coincidence with a 511~keV hit and for multiplicity~1 events.
We found an average disagreement in this ratio of 2.2\%, consistent with a relative uncertainty of 0.105 added to effects such as dead layers.
This produces our dominant systematic uncertainty contribution of $0.23\%$.

\begin{table}[]
    \centering
    \begin{tabular}{c|c|c|c|c}
        Dataset & Time & Efficiency & Exposure & BG Index\\
        & Period & & (kg-y) & (cts/keV-kg-y) \\
        \hline
        DS I M1 & 7/15-10/15 & $2.04(24)\%$ & $1.92(1)$ & $0.056(23)$ \\
        DS II M1 & 12/15-8/16  & $2.12(24)\%$ & $5.02(3)$ & $0.021(9)$ \\
        DS III M1 & 8/16-11/19 & $2.83(25)\%$ & $39.68(22)$ & $0.028(4)$ \\
        DS III M2 & 8/16-11/19 & $0.99(22)\%$ & $30.90(16)$ & $0.012(3)$ \\
        DS IV M1 & 11/19-8/20 & $2.07(25)\%$ & $7.36(4)$ & $0.015(6)$ \\
        DS V M1 & 8/20-3/21 & $2.30(24)\%$ & $7.14(4)$ & $0.023(8)$ \\
        DS V M2 & 8/20-3/21 & $3.62(27)\%$ & $6.14(4)$ & $0.018(7)$
    \end{tabular}
    \caption{Detection efficiency for \bb\ decay of $^{76}$Ge to $0^+_1$ \es\ of $^{76}$Se, isotopic exposure, and best-fit background index for each dataset (DS)/module (M).}
    \label{tab:details}
\end{table}

The measured half-life is calculated using
\begin{equation}
    T_{1/2} = \frac{\mathrm{ln}2 N_A \varepsilon M_{iso}T_{live}}{m_{76}\langle s \rangle}
\end{equation}
where $N_A$ is Avogadro's number, $m_{76}=75.9$~g is the molar mass of $^{76}$Ge, and $\langle s \rangle$ is the estimated combined amplitude of the signal peaks.
The isotopic exposure $M_{iso}T_{live}=98.2\pm0.5$~kg-y is the product of the total mass of $^{76}$Ge in a module times the operating time of the module summed over datasets.
This differs from active exposure, defined in Ref.~\cite{mjd2023} to subtract inactive detector volume and dead time; instead, these effects are included as reductions in detection efficiency, and variation in the number of active detectors drives the change in efficiency between datasets seen in Tab.~\ref{tab:details} (particularly the jump in Module~2, caused by an upgrade resulting in almost all detectors being active).

\begin{figure}
    \centering
    \includegraphics[width=0.48\textwidth]{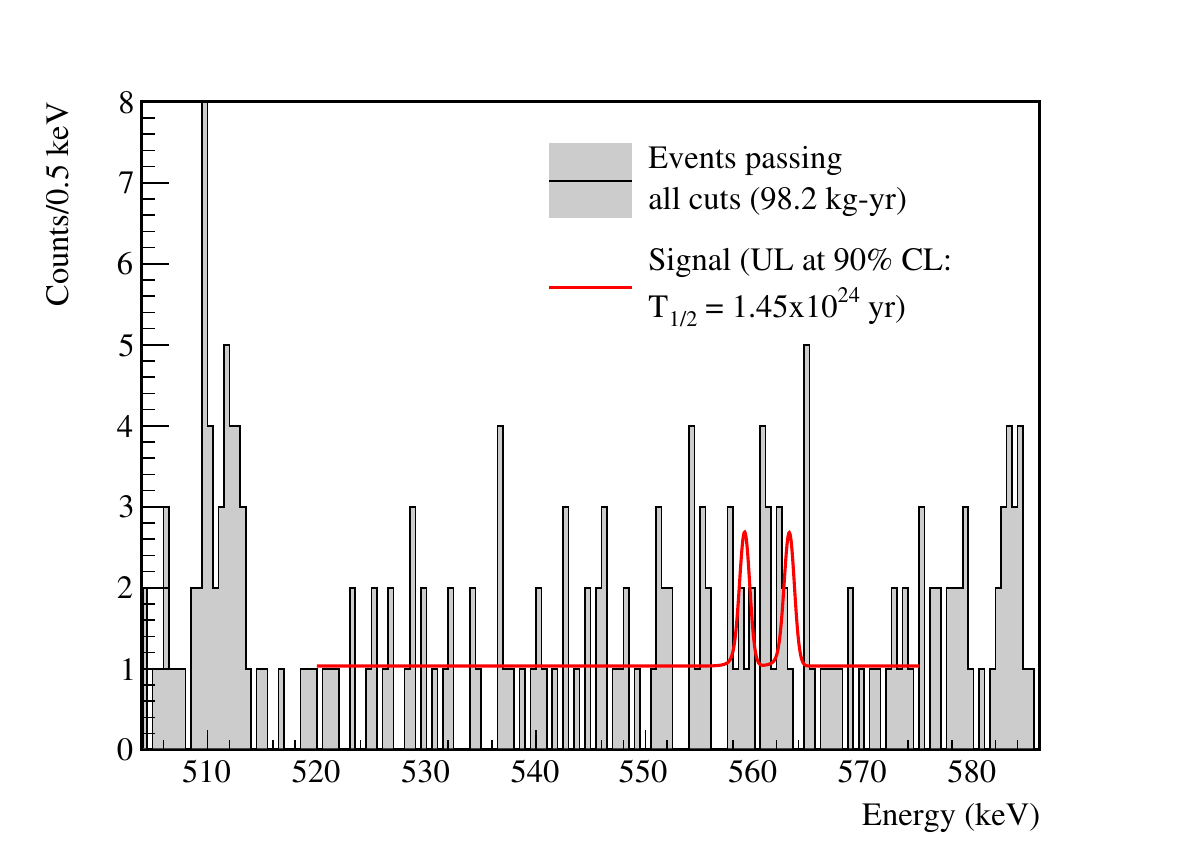}
    \caption{\label{fig:result}
        Energy spectrum for events passing all cuts for \tnbb\ to the $0^+_1$ \es. In red, the modelled signal assuming a half-life at the 90\% CL limit and the best-fit mean background.
        } 
\end{figure}

\begin{table*}[]
    \centering
    \begin{tabular}{c|c|c|c|c|c|c|c}
        Decay Mode & Peak Energies (keV) & Peak FWHM (keV) & Efficiency & $\langle s\rangle$ BF & $\langle s\rangle$ Limit & $T_{1/2}$ Limit & $T_{1/2}$ Sensitivity \\
        \hline
        $0^+_{g.s.}\xrightarrow{2\nu\beta\beta}0^+_{1}$ & 559.1, 563.2 & 1.12, 1.13 & $2.15(24)\%$ & 1.3 & 8.0 & $1.5\times10^{24}$~y & $2.2\times10^{24}$~y \\
        $0^+_{g.s.}\xrightarrow{2\nu\beta\beta}2^+_{1}$ & 559.1 & 1.12 & $1.15(26)\%$ & 0.0 & 2.1 & $3.0\times10^{24}$~y & $2.1\times10^{24}$~y \\
        $0^+_{g.s.}\xrightarrow{2\nu\beta\beta}2^+_{2}$ & 559.1, 657.0, 1216.1 & 1.12, 1.22, 1.73 & $1.76(29)\%$ & 2.1 & 10.9 & $0.88\times10^{24}$~y & $1.5\times10^{24}$~y \\
        $0^+_{g.s.}\xrightarrow{0\nu\beta\beta}0^+_{1}$ & 559.1, 563.2 & 1.12, 1.13 & $2.83(32)\%$ & 0.0 & 2.0 & $7.6\times10^{24}$~y & $5.9\times10^{24}$~y \\
        $0^+_{g.s.}\xrightarrow{0\nu\beta\beta}2^+_{1}$ & 559.1 & 1.12 & $1.58(35)\%$ & 0.0 & 2.5 &  $3.5\times10^{24}$~y & $3.5\times10^{24}$~y \\
        $0^+_{g.s.}\xrightarrow{0\nu\beta\beta}2^+_{2}$ & 559.1, 657.0, 1216.1 & 1.12, 1.22, 1.73 & $2.16(32)\%$ & 0.0 & 1.7 &  $6.6\times10^{24}$~y & $4.3\times10^{24}$~y
    \end{tabular}
    \caption{Results for each \bb-transition of $^{76}$Ge to an \es\ of $^{76}$Se. Peak FWHM and efficiency averaged over datasets, estimated counts $\langle s\rangle$ from best fit (BF) and upper limit, and half-life limit and sensitivity at 90\% C.L. are shown.  }
    \label{tab:results}
\end{table*}

We use a profile likelihood analysis to construct Neyman confidence intervals for the half-life of each \es\ decay mode.
The data are modelled using one or more peaks, using the measured peakshape function on a flat background; we also include nuisance parameters for uncertainty in the detection efficiency, drift in the peak position, and uncertainty in the peak width.
The model is applied for hits in an energy range of $520-575$~keV for the 559-~and 563-keV peaks, $620-690$~keV for the 657~keV peak, omitting $660-670$~keV to remove a U-chain peak, and $1180-1300$~keV for the 1216~keV peak, omitting $1230-1245$~keV to remove another U-chain peak.
For each dataset, we independently calculate the exposure and detection efficiency, and we float independent background indices.
We use an extended unbinned likelihood function, implemented with iminuit~\cite{iminuit}, to calculate our confidence intervals.
For the $2^+_2$ decay modes, we simultaneously profile over all three peaks.
Wilks' theorem is applied to calculate p-values for all modes except for the \znbb\ to $2^+_1$, which had zero counts in the background window; in this case, p-values were calculated through Monte-Carlo sampling.
For all \bb\ to \es\ decay modes, we measure a null result, with detailed results shown in Tab.~\ref{tab:results} and the model drawn in Fig.~\ref{fig:result}.

Combined with the measured half-life for \tnbb\ to the \gs\ of $^{76}$Se of $T_{1/2}=2.05\times 10^{21}$~yr~\cite{reine:2023}, the limit for the $0^+_1$ \es\ corresponds to a branching ratio of $\mathrm{BR}<0.0014$ (sensitivity of $\mathrm{BR}<0.0009$).
We compare this BR limit to predictions using theoretical calculations of the PSFs~\cite{Stoica2019} and NMEs, assuming the same value for $g^{eff,2\nu}_A$ applies to each daughter state, using $\mathrm{BR}=\frac{G^{2\nu}_{\es}|M^{2\nu}_{\es}|^2}{G^{2\nu}_{\gs}|M^{2\nu}_{\gs}|^2}$.
Several variants of Quasiparticle Random Phase Approximation (QRPA) have been applied~\cite{Aunola1996, Stoica1996, Toivanen1997}, with the minimum predicted BR of 0.0045 strongly disfavored with a P-value of $1\times10^{-5}$.
An Effective Theory (ET) predicted a BR of 0.0011-0.0012~\cite{Perez2018}, which is disfavored with a p-value of 0.23.
The Nuclear Shell Model (NSM) predicted a BR of 0.00068-0.00076~\cite{gerda:ES} and the Interacting Boson Model (IBM2) predicted 0.00025-0.00027~\cite{Barea2015}, both beyond the sensitivity of this search.
For \tnbb\ to the $2^+_1$ \es, BR predictions range from $4.2\times10^{-7} - 2.1\times10^{-6}$~\cite{Unlu2014, Perez2018, Kostensalo2022}, well beyond the sensitivity of this search.


The \MJD\ has set the most stringent limits to date for all \es\ decay modes in $^{76}$Ge. We achieved sensitivity to half-life values for \tnbb\ to the $0^+_1$ state of $^{76}$Se in a range predicted by recent calculations.
We have benefited from the excellent energy resolution of the experiment and from operating the detectors in vacuum.
Future experimental efforts from the LEGEND collaboration~\cite{Legend:pCDR} will use detectors operated in a liquid argon active veto, which will increase shielding between detectors (reducing efficiency) and introduce backgrounds from $^{42}$K; thus, LEGEND will likely require new analysis techniques to significantly improve on this result.

\ackfunding

\bibliography{main.bib}

\end{document}

%% file: author_list.tex
\newcommand{\ITEP}{National Research Center ``Kurchatov Institute'', Kurchatov Complex of Theoretical and Experimental Physics, Moscow, 117218 Russia}
\newcommand{\JINR}{Joint Institute for Nuclear Research, Dubna, 141980 Russia} 
\newcommand{\lbnl}{Nuclear Science Division, Lawrence Berkeley National Laboratory, Berkeley, CA 94720, USA}
\newcommand{\lanl}{Los Alamos National Laboratory, Los Alamos, NM 87545, USA}
\newcommand{\uw}{Center for Experimental Nuclear Physics and Astrophysics, and Department of Physics, University of Washington, Seattle, WA 98195, USA}
\newcommand{\unc}{Department of Physics and Astronomy, University of North Carolina, Chapel Hill, NC 27514, USA}
\newcommand{\duke}{Department of Physics, Duke University, Durham, NC 27708, USA}
\newcommand{\ncsu}{Department of Physics, North Carolina State University, Raleigh, NC 27695, USA}	
\newcommand{\ornl}{Oak Ridge National Laboratory, Oak Ridge, TN 37830, USA}
\newcommand{\ou}{Research Center for Nuclear Physics, Osaka University, Ibaraki, Osaka 567-0047, Japan}
\newcommand{\pnnl}{Pacific Northwest National Laboratory, Richland, WA 99354, USA}
\newcommand{\sdsmt}{South Dakota Mines, Rapid City, SD 57701, USA}
\newcommand{\usc}{Department of Physics and Astronomy, University of South Carolina, Columbia, SC 29208, USA}
\newcommand{\usd}{Department of Physics, University of South Dakota, Vermillion, SD 57069, USA}  
\newcommand{\ut}{Department of Physics and Astronomy, University of Tennessee, Knoxville, TN 37916, USA}
\newcommand{\tunl}{Triangle Universities Nuclear Laboratory, Durham, NC 27708, USA}
\newcommand{\williams}{Physics Department, Williams College, Williamstown, MA 01267, USA}
\newcommand{\ciemat}{Centro de Investigaciones Energ\'{e}ticas, Medioambientales y Tecnol\'{o}gicas, CIEMAT 28040, Madrid, Spain}
\newcommand{\iu}{Center for Exploration of Energy and Matter, and Department of Physics, Indiana University, Bloomington, IN 47405, USA}
\newcommand{\ucsd}{Hal\i c{\i}o\u{g}lu Data Science Institute, Department of Physics, University of California San Diego, CA 92093, USA}

\author{I.J.~Arnquist}\affiliation{\pnnl} 
\author{F.T.~Avignone~III}\affiliation{\usc}\affiliation{\ornl}
\author{A.S.~Barabash\,\orcidlink{0000-0002-5130-0922}}\affiliation{\ITEP}
\author{E.~Blalock}\affiliation{\ncsu}\affiliation{\tunl} 
\author{B.~Bos}\affiliation{\unc}\affiliation{\tunl} 
\author{M.~Busch}\affiliation{\duke}\affiliation{\tunl}	
\author{Y.-D.~Chan}\affiliation{\lbnl}
\author{J.R.~Chapman\,\orcidlink{0009-0004-9815-2981}}\affiliation{\unc}\affiliation{\tunl}
\author{C.D.~Christofferson\,\orcidlink{0009-0005-1842-9352}}\affiliation{\sdsmt} 
\author{P.-H.~Chu\,\orcidlink{0000-0003-1372-2910}}\affiliation{\lanl} 
\author{C.~Cuesta\,\orcidlink{0000-0003-1190-7233}}\affiliation{\ciemat}	
\author{J.A.~Detwiler\,\orcidlink{0000-0002-9050-4610}}\affiliation{\uw}	
\author{Yu.~Efremenko}\affiliation{\ut}\affiliation{\ornl}
\author{H.~Ejiri}\affiliation{\ou}
\author{S.R.~Elliott\,\orcidlink{0000-0001-9361-9870}}\affiliation{\lanl}
\author{N.~Fuad\,\orcidlink{0000-0002-5445-2534}}\affiliation{\iu} 
\author{G.K.~Giovanetti}\affiliation{\williams}  
\author{M.P.~Green\,\orcidlink{0000-0002-1958-8030}}\affiliation{\ncsu}\affiliation{\tunl}\affiliation{\ornl}   
\author{J.~Gruszko\,\orcidlink{0000-0002-3777-2237}}\affiliation{\unc}\affiliation{\tunl} 
\author{I.S.~Guinn\,\orcidlink{0000-0002-2424-3272}}\affiliation{\ornl} 
\author{V.E.~Guiseppe\,\orcidlink{0000-0002-0078-7101}}\affiliation{\ornl}	
\author{C.R.~Haufe}\affiliation{\unc}\affiliation{\tunl}	
\author{R.~Henning\,\orcidlink{0000-0001-8651-2960}}\affiliation{\unc}\affiliation{\tunl}
\author{D.~Hervas~Aguilar\,\orcidlink{0000-0002-9686-0659}}\altaffiliation{Present address: Technical University of Munich, 85748 Garching, Germany} \affiliation{\unc}\affiliation{\tunl} 
\author{E.W.~Hoppe\,\orcidlink{0000-0002-8171-7323}}\affiliation{\pnnl}
\author{I.~Kim\,\orcidlink{0000-0002-8394-6613}}\altaffiliation{Present address: Lawrence Livermore National Laboratory, Livermore, CA 94550, USA}\affiliation{\lanl}
\author{R.T.~Kouzes\,\orcidlink{0000-0002-6639-4140}}\affiliation{\pnnl}
\author{T.E.~Lannen~V}\affiliation{\usc} 
\author{A.~Li\,\orcidlink{0000-0002-4844-9339}}\affiliation{\ucsd} 
\author{R.~Massarczyk\,\orcidlink{0000-0001-8001-9235}}\affiliation{\lanl}		
\author{S.J.~Meijer\,\orcidlink{0000-0002-1366-0361}}\affiliation{\lanl}	
\author{T.K.~Oli\,\orcidlink{0000-0001-8857-3716}}\altaffiliation{Present address: Argonne National Laboratory, Lemont, IL 60439, USA}\affiliation{\usd}  
\author{L.S.~Paudel\,\orcidlink{0000-0003-3100-4074}}\affiliation{\usd} 
\author{W.~Pettus\,\orcidlink{0000-0003-4947-7400}}\affiliation{\iu}	
\author{A.W.P.~Poon\,\orcidlink{0000-0003-2684-6402}}\affiliation{\lbnl}
\author{D.C.~Radford}\affiliation{\ornl}
\author{A.L.~Reine\,\orcidlink{0000-0002-5900-8299}}\affiliation{\iu}	
\author{K.~Rielage\,\orcidlink{0000-0002-7392-7152}}\affiliation{\lanl}
\author{D.C.~Schaper\,\orcidlink{0000-0002-6219-650X}}\altaffiliation{Present address: Indiana Universty, Bloomington, IN 47405, USA}\affiliation{\lanl}
\author{S.J.~Schleich\,\orcidlink{0000-0003-1878-9102}}\affiliation{\iu} 
\author{D.~Tedeschi}\affiliation{\usc}		
\author{R.L.~Varner\,\orcidlink{0000-0002-0477-7488}}\affiliation{\ornl}  
\author{S.~Vasilyev}\affiliation{\JINR}	
\author{S.L.~Watkins\,\orcidlink{0000-0003-0649-1923}}\altaffiliation{Present address: Pacific Northwest National Laboratory}\affiliation{\lanl}
\author{J.F.~Wilkerson\,\orcidlink{0000-0002-0342-0217}}\affiliation{\unc}\affiliation{\tunl}\affiliation{\ornl}    
\author{C.~Wiseman\,\orcidlink{0000-0002-4232-1326}}\affiliation{\uw}		
\author{C.-H.~Yu\,\orcidlink{0000-0002-9849-842X}}\affiliation{\ornl}
\author{B.X.~Zhu}\altaffiliation{Present address: Jet Propulsion Laboratory, California Institute of Technology, Pasadena, CA 91109, USA}\affiliation{\lanl} 
\collaboration{{\sc{Majorana}} Collaboration}
\noaffiliation

\pacs{23.40-s, 23.40.Bw, 14.60.Pq, 27.50.+j}

\newcommand{\ackfunding}{This material is based upon work supported by the U.S.~Department of Energy, Office of Science, Office of Nuclear Physics under contract / award numbers DE-AC02-05CH11231, DE-AC05-00OR22725, DE-AC05-76RL0130, DE-FG02-97ER41020, DE-FG02-97ER41033, DE-FG02-97ER41041, DE-SC0012612, DE-SC0014445, DE-SC0017594, DE-SC0018060, DE-SC0022339, and LANLEM77/LANLEM78. We acknowledge support from the Particle Astrophysics Program and Nuclear Physics Program of the National Science Foundation through grant numbers MRI-0923142, PHY-1003399, PHY-1102292, PHY-1206314, PHY-1614611, PHY-13407204, PHY-1812409, PHY-1812356, PHY-2111140, and PHY-2209530. We gratefully acknowledge the support of the Laboratory Directed Research \& Development (LDRD) program at Lawrence Berkeley National Laboratory for this work. We gratefully acknowledge the support of the U.S.~Department of Energy through the Los Alamos National Laboratory LDRD Program, the Oak Ridge National Laboratory LDRD Program, and the Pacific Northwest National Laboratory LDRD Program for this work.  We gratefully acknowledge the support of the South Dakota Board of Regents Competitive Research Grant. 
We acknowledge the support of the Natural Sciences and Engineering Research Council of Canada, funding reference number SAPIN-2017-00023, and from the Canada Foundation for Innovation John R.~Evans Leaders Fund.  
We acknowledge support from the 2020/2021 L'Or\'eal-UNESCO for Women in Science Programme.
This research used resources provided by the Oak Ridge Leadership Computing Facility at Oak Ridge National Laboratory and by the National Energy Research Scientific Computing Center, a U.S.~Department of Energy Office of Science User Facility. We thank our hosts and colleagues at the Sanford Underground Research Facility for their support.}

%% file: level_diagram.tex
\usetikzlibrary{calc}
\usetikzlibrary{arrows.meta}
\usetikzlibrary{shapes.callouts}
\tikzset{
    gslevel/.style = {
        ultra thick,
        black,
    },
    eslevel/.style = {
        thick,
        black,
    },
    connect/.style = {
        dashed,
        black,
    },
}

\begin{tikzpicture}[font=\boldmath]

  \coordinate (GeLen) at (-1., 0);
  \coordinate (SeLen) at (3, 0);

  \coordinate (GeGS) at (1, 3.5);
  \coordinate (AsGS) at (1.5, 4.25);
  \coordinate (SeGS) at (3.6, 0);
  \coordinate (Se21) at (3.6, 1.25);
  \coordinate (Se01) at (3.6, 2.5);
  \coordinate (Se22) at (3.6, 3.1);
  
  \draw[gslevel] (GeGS) -- node[above, at end, xshift=8]{$0^+$} node[below] {$^{76}\mathrm{Ge}$} +(GeLen);
  \draw[gslevel] (AsGS) -- node[above, at start, xshift=8]{$2^-$} node[below] {$^{76}\mathrm{As}$} +(1.2, 0);

  \draw[eslevel] (Se22) -- node[above, at start, xshift=8] {$2^+_2$} node[at end, right] {$1216.1~\mathrm{keV}$} +(SeLen);
  \draw[eslevel] (Se01) -- node[above, at start, xshift=8] {$0^+_1$} node[at end, right] {$1122.3~\mathrm{keV}$} +(SeLen);
  \draw[eslevel] (Se21) -- node[above, at start, xshift=8] {$2^+_1$} node[at end, right] {$559.1~\mathrm{keV}$} +(SeLen);
  \draw[gslevel] (SeGS) -- node[above, at start, xshift=8] {$0^+_{g.s.}$} node[at end, right] {$0~\mathrm{keV}$} node[below] {$^{76}\mathrm{Se}$} +(SeLen);

  \draw[connect] (GeGS) edge node[sloped, below left, pos=0.9]{$Q_{\beta\beta}=2039.1~\mathrm{keV}$} (SeGS)
  (GeGS) edge node[sloped, below left, pos=1.1]{$916.8~\mathrm{keV}$} (Se01)
  (GeGS) edge node[sloped, below left, pos=1.1]{$1480.0~\mathrm{keV}$} (Se21)
  (GeGS) edge node[sloped, above, pos=0.65]{$823.0~\mathrm{keV}$} (Se22);

  \draw[-{latex[length=12pt]}] ($(Se21)+(1.2, 0)$)-- node[below, rotate=90] {$559.1$}($(SeGS)+(1.2, 0)$);
  \draw[-{latex[length=12pt]}] ($(Se01)+(0.75, 0)$)-- node[below, rotate=90] {$563.2$}($(Se21)+(0.75, 0)$);
  \draw[preaction={draw, line width=3pt, white}] [-{latex[length=12pt]}] ($(Se22)+(1.65, 0)$)-- node[below, rotate=90, xshift=-6]{$657.0$} node[above, at start]{$64\%$} ($(Se21)+(1.65, 0)$);
  \draw[preaction={draw, line width=3pt, white}] [-{latex[length=12pt]}] ($(Se22)+(2.5, 0)$)-- node[below, rotate=90, xshift=-24] {$1216.1$} node[above, at start]{$36\%$} ($(SeGS)+(2.5, 0)$);
\end{tikzpicture}

%% file: main.bbl
\begin{thebibliography}{70}%
\makeatletter
\providecommand \@ifxundefined [1]{%
 \@ifx{#1\undefined}
}%
\providecommand \@ifnum [1]{%
 \ifnum #1\expandafter \@firstoftwo
 \else \expandafter \@secondoftwo
 \fi
}%
\providecommand \@ifx [1]{%
 \ifx #1\expandafter \@firstoftwo
 \else \expandafter \@secondoftwo
 \fi
}%
\providecommand \natexlab [1]{#1}%
\providecommand \enquote  [1]{``#1''}%
\providecommand \bibnamefont  [1]{#1}%
\providecommand \bibfnamefont [1]{#1}%
\providecommand \citenamefont [1]{#1}%
\providecommand \href@noop [0]{\@secondoftwo}%
\providecommand \href [0]{\begingroup \@sanitize@url \@href}%
\providecommand \@href[1]{\@@startlink{#1}\@@href}%
\providecommand \@@href[1]{\endgroup#1\@@endlink}%
\providecommand \@sanitize@url [0]{\catcode `\\12\catcode `\$12\catcode
  `\&12\catcode `\#12\catcode `\^12\catcode `\_12\catcode `\%12\relax}%
\providecommand \@@startlink[1]{}%
\providecommand \@@endlink[0]{}%
\providecommand \url  [0]{\begingroup\@sanitize@url \@url }%
\providecommand \@url [1]{\endgroup\@href {#1}{\urlprefix }}%
\providecommand \urlprefix  [0]{URL }%
\providecommand \Eprint [0]{\href }%
\providecommand \doibase [0]{https://doi.org/}%
\providecommand \selectlanguage [0]{\@gobble}%
\providecommand \bibinfo  [0]{\@secondoftwo}%
\providecommand \bibfield  [0]{\@secondoftwo}%
\providecommand \translation [1]{[#1]}%
\providecommand \BibitemOpen [0]{}%
\providecommand \bibitemStop [0]{}%
\providecommand \bibitemNoStop [0]{.\EOS\space}%
\providecommand \EOS [0]{\spacefactor3000\relax}%
\providecommand \BibitemShut  [1]{\csname bibitem#1\endcsname}%
\let\auto@bib@innerbib\@empty
\bibitem [{\citenamefont {Goeppert-Mayer}(1935)}]{GoeppertMayer1935}%
  \BibitemOpen
  \bibfield  {author} {\bibinfo {author} {\bibfnamefont {M.}~\bibnamefont
  {Goeppert-Mayer}},\ }\bibfield  {title} {\bibinfo {title} {Double
  beta-disintegration},\ }\href {https://doi.org/10.1103/PhysRev.48.512}
  {\bibfield  {journal} {\bibinfo  {journal} {Phys. Rev.}\ }\textbf {\bibinfo
  {volume} {48}},\ \bibinfo {pages} {512} (\bibinfo {year} {1935})}\BibitemShut
  {NoStop}%
\bibitem [{\citenamefont {Majorana}(2008)}]{Majorana1937}%
  \BibitemOpen
  \bibfield  {author} {\bibinfo {author} {\bibfnamefont {E.}~\bibnamefont
  {Majorana}},\ }\bibfield  {title} {\bibinfo {title} {Teoria simmetrica
  dell'elettrone e del positrone},\ }\href {https://doi.org/10.1007/BF02961314}
  {\bibfield  {journal} {\bibinfo  {journal} {Il Nuovo Cimento (1924-1942)}\
  }\textbf {\bibinfo {volume} {14}},\ \bibinfo {pages} {171} (\bibinfo {year}
  {2008})}\BibitemShut {NoStop}%
\bibitem [{\citenamefont {Furry}(1939)}]{Furry1939}%
  \BibitemOpen
  \bibfield  {author} {\bibinfo {author} {\bibfnamefont {W.~H.}\ \bibnamefont
  {Furry}},\ }\bibfield  {title} {\bibinfo {title} {On transition probabilities
  in double beta-disintegration},\ }\href
  {https://doi.org/10.1103/PhysRev.56.1184} {\bibfield  {journal} {\bibinfo
  {journal} {Phys. Rev.}\ }\textbf {\bibinfo {volume} {56}},\ \bibinfo {pages}
  {1184} (\bibinfo {year} {1939})}\BibitemShut {NoStop}%
\bibitem [{\citenamefont {Barabash}(2020)}]{Barabash2020}%
  \BibitemOpen
  \bibfield  {author} {\bibinfo {author} {\bibfnamefont {A.}~\bibnamefont
  {Barabash}},\ }\bibfield  {title} {\bibinfo {title} {Precise half-life values
  for two-neutrino double-$\beta$ decay: 2020 review},\ }\href
  {https://doi.org/10.3390/universe6100159} {\bibfield  {journal} {\bibinfo
  {journal} {Universe}\ }\textbf {\bibinfo {volume} {6}},\ \bibinfo {pages}
  {159} (\bibinfo {year} {2020})}\BibitemShut {NoStop}%
\bibitem [{\citenamefont {Schechter}\ and\ \citenamefont
  {Valle}(1982)}]{Schechter1982}%
  \BibitemOpen
  \bibfield  {author} {\bibinfo {author} {\bibfnamefont {J.}~\bibnamefont
  {Schechter}}\ and\ \bibinfo {author} {\bibfnamefont {J.~W.~F.}\ \bibnamefont
  {Valle}},\ }\bibfield  {title} {\bibinfo {title} {Neutrinoless
  double-$\ensuremath{\beta}$ decay in {SU(2)\texttimes U(1)} theories},\
  }\href {https://doi.org/10.1103/PhysRevD.25.2951} {\bibfield  {journal}
  {\bibinfo  {journal} {Phys. Rev. D}\ }\textbf {\bibinfo {volume} {25}},\
  \bibinfo {pages} {2951} (\bibinfo {year} {1982})}\BibitemShut {NoStop}%
\bibitem [{\citenamefont {{Sakharov}}(1967)}]{Sakharov1967}%
  \BibitemOpen
  \bibfield  {author} {\bibinfo {author} {\bibfnamefont {A.~D.}\ \bibnamefont
  {{Sakharov}}},\ }\bibfield  {title} {\bibinfo {title} {{Violation of CP
  Invariance, C Asymmetry, and Baryon Asymmetry of the Universe}},\ }\href
  {https://doi.org/10.1070/PU1991v034n05ABEH002497} {\bibfield  {journal}
  {\bibinfo  {journal} {Pisma Zh. Eksp. Teor. Fiz.}\ }\textbf {\bibinfo
  {volume} {5}},\ \bibinfo {pages} {32} (\bibinfo {year} {1967})}\BibitemShut
  {NoStop}%
\bibitem [{\citenamefont {Fukugita}\ and\ \citenamefont
  {Yanagida}(1986)}]{Fukugita1986}%
  \BibitemOpen
  \bibfield  {author} {\bibinfo {author} {\bibfnamefont {M.}~\bibnamefont
  {Fukugita}}\ and\ \bibinfo {author} {\bibfnamefont {T.}~\bibnamefont
  {Yanagida}},\ }\bibfield  {title} {\bibinfo {title} {Baryogenesis without
  grand unification},\ }\href
  {https://doi.org/https://doi.org/10.1016/0370-2693(86)91126-3} {\bibfield
  {journal} {\bibinfo  {journal} {Phys. Lett. B}\ }\textbf {\bibinfo {volume}
  {174}},\ \bibinfo {pages} {45 } (\bibinfo {year} {1986})}\BibitemShut
  {NoStop}%
\bibitem [{\citenamefont {Dolinski}\ \emph {et~al.}(2019)\citenamefont
  {Dolinski}, \citenamefont {Poon},\ and\ \citenamefont
  {Rodejohann}}]{Dolinski2019}%
  \BibitemOpen
  \bibfield  {author} {\bibinfo {author} {\bibfnamefont {M.~J.}\ \bibnamefont
  {Dolinski}}, \bibinfo {author} {\bibfnamefont {A.~W.}\ \bibnamefont {Poon}},\
  and\ \bibinfo {author} {\bibfnamefont {W.}~\bibnamefont {Rodejohann}},\
  }\bibfield  {title} {\bibinfo {title} {Neutrinoless double-beta decay: Status
  and prospects},\ }\href
  {https://doi.org/https://doi.org/10.1146/annurev-nucl-101918-023407}
  {\bibfield  {journal} {\bibinfo  {journal} {Annual Review of Nuclear and
  Particle Science}\ }\textbf {\bibinfo {volume} {69}},\ \bibinfo {pages} {219}
  (\bibinfo {year} {2019})}\BibitemShut {NoStop}%
\bibitem [{\citenamefont {Gómez-Cadenas}\ \emph {et~al.}(2023)\citenamefont
  {Gómez-Cadenas}, \citenamefont {Martín-Albo}, \citenamefont {Menéndez},
  \citenamefont {Mezzetto}, \citenamefont {Monrabal},\ and\ \citenamefont
  {Sorel}}]{Gomez-Cadenas2023}%
  \BibitemOpen
  \bibfield  {author} {\bibinfo {author} {\bibfnamefont {J.~J.}\ \bibnamefont
  {Gómez-Cadenas}}, \bibinfo {author} {\bibfnamefont {J.}~\bibnamefont
  {Martín-Albo}}, \bibinfo {author} {\bibfnamefont {J.}~\bibnamefont
  {Menéndez}}, \bibinfo {author} {\bibfnamefont {M.}~\bibnamefont {Mezzetto}},
  \bibinfo {author} {\bibfnamefont {F.}~\bibnamefont {Monrabal}},\ and\
  \bibinfo {author} {\bibfnamefont {M.}~\bibnamefont {Sorel}},\ }\bibfield
  {title} {\bibinfo {title} {The search for neutrinoless double-beta decay},\
  }\href {https://doi.org/10.1007/s40766-023-00049-2} {\bibfield  {journal}
  {\bibinfo  {journal} {Riv. Nuovo Cimento}\ }\textbf {\bibinfo {volume}
  {46}},\ \bibinfo {pages} {619} (\bibinfo {year} {2023})}\BibitemShut
  {NoStop}%
\bibitem [{\citenamefont {Barabash}(2023)}]{Barabash2023}%
  \BibitemOpen
  \bibfield  {author} {\bibinfo {author} {\bibfnamefont {A.}~\bibnamefont
  {Barabash}},\ }\bibfield  {title} {\bibinfo {title} {Double beta decay
  experiments: Recent achievements and future prospects},\ }\href
  {https://doi.org/10.3390/universe9060290} {\bibfield  {journal} {\bibinfo
  {journal} {Universe}\ }\textbf {\bibinfo {volume} {9}},\ \bibinfo {pages}
  {290} (\bibinfo {year} {2023})}\BibitemShut {NoStop}%
\bibitem [{\citenamefont {Agostini}\ \emph {et~al.}(2023)\citenamefont
  {Agostini}, \citenamefont {Benato}, \citenamefont {Detwiler}, \citenamefont
  {Men\'endez},\ and\ \citenamefont {Vissani}}]{Agostini2023}%
  \BibitemOpen
  \bibfield  {author} {\bibinfo {author} {\bibfnamefont {M.}~\bibnamefont
  {Agostini}}, \bibinfo {author} {\bibfnamefont {G.}~\bibnamefont {Benato}},
  \bibinfo {author} {\bibfnamefont {J.~A.}\ \bibnamefont {Detwiler}}, \bibinfo
  {author} {\bibfnamefont {J.}~\bibnamefont {Men\'endez}},\ and\ \bibinfo
  {author} {\bibfnamefont {F.}~\bibnamefont {Vissani}},\ }\bibfield  {title}
  {\bibinfo {title} {Toward the discovery of matter creation with neutrinoless
  $\ensuremath{\beta}\ensuremath{\beta}$ decay},\ }\href
  {https://doi.org/10.1103/RevModPhys.95.025002} {\bibfield  {journal}
  {\bibinfo  {journal} {Rev. Mod. Phys.}\ }\textbf {\bibinfo {volume} {95}},\
  \bibinfo {pages} {025002} (\bibinfo {year} {2023})}\BibitemShut {NoStop}%
\bibitem [{\citenamefont {Belli}\ \emph {et~al.}(2020)\citenamefont {Belli},
  \citenamefont {Bernabei}, \citenamefont {Cappella}, \citenamefont
  {Caracciolo}, \citenamefont {Cerulli}, \citenamefont {Incicchitti},\ and\
  \citenamefont {Merlo}}]{Belli2020}%
  \BibitemOpen
  \bibfield  {author} {\bibinfo {author} {\bibfnamefont {P.}~\bibnamefont
  {Belli}}, \bibinfo {author} {\bibfnamefont {R.}~\bibnamefont {Bernabei}},
  \bibinfo {author} {\bibfnamefont {F.}~\bibnamefont {Cappella}}, \bibinfo
  {author} {\bibfnamefont {V.}~\bibnamefont {Caracciolo}}, \bibinfo {author}
  {\bibfnamefont {R.}~\bibnamefont {Cerulli}}, \bibinfo {author} {\bibfnamefont
  {A.}~\bibnamefont {Incicchitti}},\ and\ \bibinfo {author} {\bibfnamefont
  {V.}~\bibnamefont {Merlo}},\ }\bibfield  {title} {\bibinfo {title} {Double
  beta decay to excited states of daughter nuclei},\ }\href
  {https://doi.org/10.3390/universe6120239} {\bibfield  {journal} {\bibinfo
  {journal} {Universe}\ }\textbf {\bibinfo {volume} {6}},\ \bibinfo {pages}
  {239} (\bibinfo {year} {2020})}\BibitemShut {NoStop}%
\bibitem [{\citenamefont {Barabash}\ \emph {et~al.}(1995)\citenamefont
  {Barabash}, \citenamefont {Avignone}, \citenamefont {Collar}, \citenamefont
  {Guerard}, \citenamefont {Arthur}, \citenamefont {Brodzinski}, \citenamefont
  {Miley}, \citenamefont {Reeves}, \citenamefont {Meier}, \citenamefont
  {Ruddick},\ and\ \citenamefont {Umatov}}]{Barabash1995}%
  \BibitemOpen
  \bibfield  {author} {\bibinfo {author} {\bibfnamefont {A.}~\bibnamefont
  {Barabash}}, \bibinfo {author} {\bibfnamefont {F.}~\bibnamefont {Avignone}},
  \bibinfo {author} {\bibfnamefont {J.}~\bibnamefont {Collar}}, \bibinfo
  {author} {\bibfnamefont {C.}~\bibnamefont {Guerard}}, \bibinfo {author}
  {\bibfnamefont {R.}~\bibnamefont {Arthur}}, \bibinfo {author} {\bibfnamefont
  {R.}~\bibnamefont {Brodzinski}}, \bibinfo {author} {\bibfnamefont
  {H.}~\bibnamefont {Miley}}, \bibinfo {author} {\bibfnamefont
  {J.}~\bibnamefont {Reeves}}, \bibinfo {author} {\bibfnamefont
  {J.}~\bibnamefont {Meier}}, \bibinfo {author} {\bibfnamefont
  {K.}~\bibnamefont {Ruddick}},\ and\ \bibinfo {author} {\bibfnamefont
  {V.}~\bibnamefont {Umatov}},\ }\bibfield  {title} {\bibinfo {title} {Two
  neutrino double-beta decay of {$^{100}$Mo} to the first excited $0^+$ state
  in {$^{100}$Ru}},\ }\href
  {https://doi.org/https://doi.org/10.1016/0370-2693(94)01657-X} {\bibfield
  {journal} {\bibinfo  {journal} {Physics Letters B}\ }\textbf {\bibinfo
  {volume} {345}},\ \bibinfo {pages} {408 } (\bibinfo {year}
  {1995})}\BibitemShut {NoStop}%
\bibitem [{\citenamefont {Barabash}\ \emph {et~al.}(1999)\citenamefont
  {Barabash}, \citenamefont {Gurriaran}, \citenamefont {Hubert}, \citenamefont
  {Hubert},\ and\ \citenamefont {Umatov}}]{Barabash1999}%
  \BibitemOpen
  \bibfield  {author} {\bibinfo {author} {\bibfnamefont {A.}~\bibnamefont
  {Barabash}}, \bibinfo {author} {\bibfnamefont {R.}~\bibnamefont {Gurriaran}},
  \bibinfo {author} {\bibfnamefont {F.}~\bibnamefont {Hubert}}, \bibinfo
  {author} {\bibfnamefont {P.}~\bibnamefont {Hubert}},\ and\ \bibinfo {author}
  {\bibfnamefont {V.}~\bibnamefont {Umatov}},\ }\bibfield  {title} {\bibinfo
  {title} {$2\nu\beta\beta$ decay of {$^{100}$Mo} to the first $0^+$ excited
  state in {$^{100}$Ru}},\ }\href
  {https://doi.org/http://dx.doi.org/10.1134/1.855586} {\bibfield  {journal}
  {\bibinfo  {journal} {Phys. At. Nucl.}\ }\textbf {\bibinfo {volume} {62}},\
  \bibinfo {pages} {2039} (\bibinfo {year} {1999})}\BibitemShut {NoStop}%
\bibitem [{\citenamefont {De~Braeckeleer}\ \emph {et~al.}(2001)\citenamefont
  {De~Braeckeleer}, \citenamefont {Hornish}, \citenamefont {Barabash},\ and\
  \citenamefont {Umatov}}]{DeBraeckeleer2001}%
  \BibitemOpen
  \bibfield  {author} {\bibinfo {author} {\bibfnamefont {L.}~\bibnamefont
  {De~Braeckeleer}}, \bibinfo {author} {\bibfnamefont {M.}~\bibnamefont
  {Hornish}}, \bibinfo {author} {\bibfnamefont {A.}~\bibnamefont {Barabash}},\
  and\ \bibinfo {author} {\bibfnamefont {V.}~\bibnamefont {Umatov}},\
  }\bibfield  {title} {\bibinfo {title} {Measurement of the
  $\mathit{\ensuremath{\beta}}\mathit{\ensuremath{\beta}}$-decay rate of
  ${}^{100}\mathrm{Mo}$ to the first excited ${0}^{+}$ state of
  ${}^{100}\mathrm{Ru}$},\ }\href {https://doi.org/10.1103/PhysRevLett.86.3510}
  {\bibfield  {journal} {\bibinfo  {journal} {Phys. Rev. Lett.}\ }\textbf
  {\bibinfo {volume} {86}},\ \bibinfo {pages} {3510} (\bibinfo {year}
  {2001})}\BibitemShut {NoStop}%
\bibitem [{\citenamefont {Arnold}\ \emph {et~al.}(2007)\citenamefont {Arnold}
  \emph {et~al.}}]{Arnold2007}%
  \BibitemOpen
  \bibfield  {author} {\bibinfo {author} {\bibfnamefont {R.}~\bibnamefont
  {Arnold}} \emph {et~al.},\ }\bibfield  {title} {\bibinfo {title} {Measurement
  of double beta decay of {$^{100}$Mo} to excited states in the {NEMO 3}
  experiment},\ }\href
  {https://doi.org/https://doi.org/10.1016/j.nuclphysa.2006.09.021} {\bibfield
  {journal} {\bibinfo  {journal} {Nuclear Physics A}\ }\textbf {\bibinfo
  {volume} {781}},\ \bibinfo {pages} {209} (\bibinfo {year}
  {2007})}\BibitemShut {NoStop}%
\bibitem [{\citenamefont {Kidd}\ \emph {et~al.}(2009)\citenamefont {Kidd},
  \citenamefont {Esterline}, \citenamefont {Tornow}, \citenamefont {Barabash},\
  and\ \citenamefont {Umatov}}]{Kidd2009}%
  \BibitemOpen
  \bibfield  {author} {\bibinfo {author} {\bibfnamefont {M.}~\bibnamefont
  {Kidd}}, \bibinfo {author} {\bibfnamefont {J.}~\bibnamefont {Esterline}},
  \bibinfo {author} {\bibfnamefont {W.}~\bibnamefont {Tornow}}, \bibinfo
  {author} {\bibfnamefont {A.}~\bibnamefont {Barabash}},\ and\ \bibinfo
  {author} {\bibfnamefont {V.}~\bibnamefont {Umatov}},\ }\bibfield  {title}
  {\bibinfo {title} {New results for double-beta decay of {$^{100}$Mo} to
  excited final states of {$^{100}$Ru} using the {TUNL-ITEP} apparatus},\
  }\href {https://doi.org/https://doi.org/10.1016/j.nuclphysa.2009.01.082}
  {\bibfield  {journal} {\bibinfo  {journal} {Nuclear Physics A}\ }\textbf
  {\bibinfo {volume} {821}},\ \bibinfo {pages} {251} (\bibinfo {year}
  {2009})}\BibitemShut {NoStop}%
\bibitem [{\citenamefont {Belli}\ \emph {et~al.}(2010)\citenamefont {Belli}
  \emph {et~al.}}]{Belli2014}%
  \BibitemOpen
  \bibfield  {author} {\bibinfo {author} {\bibfnamefont {P.}~\bibnamefont
  {Belli}} \emph {et~al.},\ }\bibfield  {title} {\bibinfo {title} {New
  observation of $2\beta2\nu$ decay of {$^{100}$Mo} to the $0_1^+$ level of
  {$^{100}$Ru} in the {ARMONIA} experiment},\ }\href
  {https://doi.org/https://doi.org/10.1016/j.nuclphysa.2010.06.010} {\bibfield
  {journal} {\bibinfo  {journal} {Nuclear Physics A}\ }\textbf {\bibinfo
  {volume} {846}},\ \bibinfo {pages} {143} (\bibinfo {year}
  {2010})}\BibitemShut {NoStop}%
\bibitem [{\citenamefont {Arnold}\ \emph {et~al.}(2014)\citenamefont {Arnold}
  \emph {et~al.}}]{Arnold2014}%
  \BibitemOpen
  \bibfield  {author} {\bibinfo {author} {\bibfnamefont {R.}~\bibnamefont
  {Arnold}} \emph {et~al.},\ }\bibfield  {title} {\bibinfo {title}
  {Investigation of double beta decay of {$^{100}$Mo} to excited states of
  {$^{100}$Ru}},\ }\href
  {https://doi.org/https://doi.org/10.1016/j.nuclphysa.2014.01.008} {\bibfield
  {journal} {\bibinfo  {journal} {Nuclear Physics A}\ }\textbf {\bibinfo
  {volume} {925}},\ \bibinfo {pages} {25} (\bibinfo {year} {2014})}\BibitemShut
  {NoStop}%
\bibitem [{\citenamefont {Augier}\ \emph {et~al.}(2023)\citenamefont {Augier}
  \emph {et~al.}}]{Augier2023}%
  \BibitemOpen
  \bibfield  {author} {\bibinfo {author} {\bibfnamefont {C.}~\bibnamefont
  {Augier}} \emph {et~al.} (\bibinfo {collaboration} {CUPID-Mo
  Collaboration}),\ }\bibfield  {title} {\bibinfo {title} {New measurement of
  double-$\ensuremath{\beta}$ decays of $^{100}\mathrm{Mo}$ to excited states
  of $^{100}\mathrm{Ru}$ with the cupid-mo experiment},\ }\href
  {https://doi.org/10.1103/PhysRevC.107.025503} {\bibfield  {journal} {\bibinfo
   {journal} {Phys. Rev. C}\ }\textbf {\bibinfo {volume} {107}},\ \bibinfo
  {pages} {025503} (\bibinfo {year} {2023})}\BibitemShut {NoStop}%
\bibitem [{\citenamefont {Barabash}\ \emph {et~al.}(2004)\citenamefont
  {Barabash}, \citenamefont {Hubert}, \citenamefont {Hubert},\ and\
  \citenamefont {Umatov}}]{Barabash2004}%
  \BibitemOpen
  \bibfield  {author} {\bibinfo {author} {\bibfnamefont {A.~S.}\ \bibnamefont
  {Barabash}}, \bibinfo {author} {\bibfnamefont {F.}~\bibnamefont {Hubert}},
  \bibinfo {author} {\bibfnamefont {P.}~\bibnamefont {Hubert}},\ and\ \bibinfo
  {author} {\bibfnamefont {V.~I.}\ \bibnamefont {Umatov}},\ }\bibfield  {title}
  {\bibinfo {title} {Double beta decay of {$^{150}$Nd} to the first $0^+$
  excited state of {$^{150}$Sm}},\ }\href
  {https://doi.org/https://doi.org/10.1134/1.1675911} {\bibfield  {journal}
  {\bibinfo  {journal} {JETP Lett.}\ }\textbf {\bibinfo {volume} {79}},\
  \bibinfo {pages} {10} (\bibinfo {year} {2004})}\BibitemShut {NoStop}%
\bibitem [{\citenamefont {Barabash}\ \emph {et~al.}(2009)\citenamefont
  {Barabash}, \citenamefont {Hubert}, \citenamefont {Nachab},\ and\
  \citenamefont {Umatov}}]{Barabash2009}%
  \BibitemOpen
  \bibfield  {author} {\bibinfo {author} {\bibfnamefont {A.~S.}\ \bibnamefont
  {Barabash}}, \bibinfo {author} {\bibfnamefont {P.}~\bibnamefont {Hubert}},
  \bibinfo {author} {\bibfnamefont {A.}~\bibnamefont {Nachab}},\ and\ \bibinfo
  {author} {\bibfnamefont {V.~I.}\ \bibnamefont {Umatov}},\ }\bibfield  {title}
  {\bibinfo {title} {Investigation of $\ensuremath{\beta}\ensuremath{\beta}$
  decay in $^{150}\mathrm{Nd}$ and $^{148}\mathrm{Nd}$ to the excited states of
  daughter nuclei},\ }\href {https://doi.org/10.1103/PhysRevC.79.045501}
  {\bibfield  {journal} {\bibinfo  {journal} {Phys. Rev. C}\ }\textbf {\bibinfo
  {volume} {79}},\ \bibinfo {pages} {045501} (\bibinfo {year}
  {2009})}\BibitemShut {NoStop}%
\bibitem [{\citenamefont {Kidd}\ \emph {et~al.}(2014)\citenamefont {Kidd},
  \citenamefont {Esterline}, \citenamefont {Finch},\ and\ \citenamefont
  {Tornow}}]{Kidd2014}%
  \BibitemOpen
  \bibfield  {author} {\bibinfo {author} {\bibfnamefont {M.~F.}\ \bibnamefont
  {Kidd}}, \bibinfo {author} {\bibfnamefont {J.~H.}\ \bibnamefont {Esterline}},
  \bibinfo {author} {\bibfnamefont {S.~W.}\ \bibnamefont {Finch}},\ and\
  \bibinfo {author} {\bibfnamefont {W.}~\bibnamefont {Tornow}},\ }\bibfield
  {title} {\bibinfo {title} {Two-neutrino double-$\ensuremath{\beta}$ decay of
  $^{150}\mathrm{Nd}$ to excited final states in $^{150}\mathrm{Sm}$},\ }\href
  {https://doi.org/10.1103/PhysRevC.90.055501} {\bibfield  {journal} {\bibinfo
  {journal} {Phys. Rev. C}\ }\textbf {\bibinfo {volume} {90}},\ \bibinfo
  {pages} {055501} (\bibinfo {year} {2014})}\BibitemShut {NoStop}%
\bibitem [{\citenamefont {Polischuk}\ \emph {et~al.}(2021)\citenamefont
  {Polischuk}, \citenamefont {Barabash}, \citenamefont {Belli}, \citenamefont
  {Bernabei}, \citenamefont {Boiko}, \citenamefont {Cappella}, \citenamefont
  {Caracciolo}, \citenamefont {Cerulli}, \citenamefont {Danevich},
  \citenamefont {Marco}, \citenamefont {Incicchitti}, \citenamefont
  {Kasperovych}, \citenamefont {Kobychev}, \citenamefont {Konovalov},
  \citenamefont {Laubenstein}, \citenamefont {Poda}, \citenamefont {Tretyak},\
  and\ \citenamefont {Umatov}}]{Polischuk2021}%
  \BibitemOpen
  \bibfield  {author} {\bibinfo {author} {\bibfnamefont {O.~G.}\ \bibnamefont
  {Polischuk}}, \bibinfo {author} {\bibfnamefont {A.~S.}\ \bibnamefont
  {Barabash}}, \bibinfo {author} {\bibfnamefont {P.}~\bibnamefont {Belli}},
  \bibinfo {author} {\bibfnamefont {R.}~\bibnamefont {Bernabei}}, \bibinfo
  {author} {\bibfnamefont {R.~S.}\ \bibnamefont {Boiko}}, \bibinfo {author}
  {\bibfnamefont {F.}~\bibnamefont {Cappella}}, \bibinfo {author}
  {\bibfnamefont {V.}~\bibnamefont {Caracciolo}}, \bibinfo {author}
  {\bibfnamefont {R.}~\bibnamefont {Cerulli}}, \bibinfo {author} {\bibfnamefont
  {F.~A.}\ \bibnamefont {Danevich}}, \bibinfo {author} {\bibfnamefont {A.~D.}\
  \bibnamefont {Marco}}, \bibinfo {author} {\bibfnamefont {A.}~\bibnamefont
  {Incicchitti}}, \bibinfo {author} {\bibfnamefont {D.~V.}\ \bibnamefont
  {Kasperovych}}, \bibinfo {author} {\bibfnamefont {V.~V.}\ \bibnamefont
  {Kobychev}}, \bibinfo {author} {\bibfnamefont {S.~I.}\ \bibnamefont
  {Konovalov}}, \bibinfo {author} {\bibfnamefont {M.}~\bibnamefont
  {Laubenstein}}, \bibinfo {author} {\bibfnamefont {D.~V.}\ \bibnamefont
  {Poda}}, \bibinfo {author} {\bibfnamefont {V.~I.}\ \bibnamefont {Tretyak}},\
  and\ \bibinfo {author} {\bibfnamefont {V.~I.}\ \bibnamefont {Umatov}},\
  }\bibfield  {title} {\bibinfo {title} {Double beta decay of {$^{150}Nd$} to
  the first $0^+$ excited level of {$^{150}$Sm}},\ }\href
  {https://doi.org/10.1088/1402-4896/ac00a5} {\bibfield  {journal} {\bibinfo
  {journal} {Physica Scripta}\ }\textbf {\bibinfo {volume} {96}},\ \bibinfo
  {pages} {085302} (\bibinfo {year} {2021})}\BibitemShut {NoStop}%
\bibitem [{\citenamefont {Aguerre}\ \emph {et~al.}(2023)\citenamefont {Aguerre}
  \emph {et~al.}}]{Aguerre2023}%
  \BibitemOpen
  \bibfield  {author} {\bibinfo {author} {\bibfnamefont {X.}~\bibnamefont
  {Aguerre}} \emph {et~al.},\ }\bibfield  {title} {\bibinfo {title}
  {Measurement of the double-$\beta$ decay of {$^{150}$Nd} to the {$0^+_1$}
  excited state of {$^{150}$Sm} in {NEMO-3}},\ }\bibfield  {journal} {\bibinfo
  {journal} {Eur. Phys. J. C}\ }\textbf {\bibinfo {volume} {83}},\ \href
  {https://doi.org/10.1140/epjc/s10052-023-12227-x}
  {10.1140/epjc/s10052-023-12227-x} (\bibinfo {year} {2023})\BibitemShut
  {NoStop}%
\bibitem [{\citenamefont {Kotila}\ and\ \citenamefont
  {Iachello}(2012)}]{Kotila2012}%
  \BibitemOpen
  \bibfield  {author} {\bibinfo {author} {\bibfnamefont {J.}~\bibnamefont
  {Kotila}}\ and\ \bibinfo {author} {\bibfnamefont {F.}~\bibnamefont
  {Iachello}},\ }\bibfield  {title} {\bibinfo {title} {Phase-space factors for
  double-$\ensuremath{\beta}$ decay},\ }\href
  {https://doi.org/10.1103/PhysRevC.85.034316} {\bibfield  {journal} {\bibinfo
  {journal} {Phys. Rev. C}\ }\textbf {\bibinfo {volume} {85}},\ \bibinfo
  {pages} {034316} (\bibinfo {year} {2012})}\BibitemShut {NoStop}%
\bibitem [{\citenamefont {Neacsu}\ and\ \citenamefont
  {Horoi}(2016)}]{Neacsu2016}%
  \BibitemOpen
  \bibfield  {author} {\bibinfo {author} {\bibfnamefont {A.}~\bibnamefont
  {Neacsu}}\ and\ \bibinfo {author} {\bibfnamefont {M.}~\bibnamefont {Horoi}},\
  }\bibfield  {title} {\bibinfo {title} {An effective method to accurately
  calculate the phase space factors for $\beta^-\beta^-$ decay},\ }\href
  {https://doi.org/https://doi.org/10.1155/2016/7486712} {\bibfield  {journal}
  {\bibinfo  {journal} {Advances in High Energy Physics}\ }\textbf {\bibinfo
  {volume} {2016}},\ \bibinfo {pages} {7486712} (\bibinfo {year}
  {2016})}\BibitemShut {NoStop}%
\bibitem [{\citenamefont {Stoica}\ and\ \citenamefont
  {Mirea}(2019)}]{Stoica2019}%
  \BibitemOpen
  \bibfield  {author} {\bibinfo {author} {\bibfnamefont {S.}~\bibnamefont
  {Stoica}}\ and\ \bibinfo {author} {\bibfnamefont {M.}~\bibnamefont {Mirea}},\
  }\bibfield  {title} {\bibinfo {title} {Phase space factors for double-beta
  decays},\ }\href {https://doi.org/10.3389/fphy.2019.00012} {\bibfield
  {journal} {\bibinfo  {journal} {Frontiers in Physics}\ }\textbf {\bibinfo
  {volume} {7}},\ \bibinfo {pages} {12} (\bibinfo {year} {2019})}\BibitemShut
  {NoStop}%
\bibitem [{\citenamefont {Suhonen}(2017)}]{Suhonen2017}%
  \BibitemOpen
  \bibfield  {author} {\bibinfo {author} {\bibfnamefont {J.~T.}\ \bibnamefont
  {Suhonen}},\ }\bibfield  {title} {\bibinfo {title} {Value of the axial-vector
  coupling strength in $\beta$ and $\beta\beta$ decays: A review},\ }\bibfield
  {journal} {\bibinfo  {journal} {Front. Phys.}\ }\textbf {\bibinfo {volume}
  {5}},\ \href {https://doi.org/10.3389/fphy.2017.00055}
  {10.3389/fphy.2017.00055} (\bibinfo {year} {2017}),\ \bibinfo {note} {cited
  by: 148; All Open Access, Gold Open Access, Green Open Access}\BibitemShut
  {NoStop}%
\bibitem [{\citenamefont {Dolgov}\ and\ \citenamefont
  {Smirnov}(2005)}]{Dolgov2005}%
  \BibitemOpen
  \bibfield  {author} {\bibinfo {author} {\bibfnamefont {A.}~\bibnamefont
  {Dolgov}}\ and\ \bibinfo {author} {\bibfnamefont {A.}~\bibnamefont
  {Smirnov}},\ }\bibfield  {title} {\bibinfo {title} {Possible violation of the
  spin-statistics relation for neutrinos: Cosmological and astrophysical
  consequences},\ }\href
  {https://doi.org/https://doi.org/10.1016/j.physletb.2005.06.035} {\bibfield
  {journal} {\bibinfo  {journal} {Phys. Lett. B}\ }\textbf {\bibinfo {volume}
  {621}},\ \bibinfo {pages} {1 } (\bibinfo {year} {2005})}\BibitemShut
  {NoStop}%
\bibitem [{\citenamefont {Barabash}\ \emph {et~al.}(2007)\citenamefont
  {Barabash}, \citenamefont {Dolgov}, \citenamefont {Dvornický}, \citenamefont
  {Šimkovic},\ and\ \citenamefont {Smirnov}}]{Barabash2007}%
  \BibitemOpen
  \bibfield  {author} {\bibinfo {author} {\bibfnamefont {A.}~\bibnamefont
  {Barabash}}, \bibinfo {author} {\bibfnamefont {A.}~\bibnamefont {Dolgov}},
  \bibinfo {author} {\bibfnamefont {R.}~\bibnamefont {Dvornický}}, \bibinfo
  {author} {\bibfnamefont {F.}~\bibnamefont {Šimkovic}},\ and\ \bibinfo
  {author} {\bibfnamefont {A.}~\bibnamefont {Smirnov}},\ }\bibfield  {title}
  {\bibinfo {title} {Statistics of neutrinos and the double beta decay},\
  }\href {https://doi.org/https://doi.org/10.1016/j.nuclphysb.2007.05.033}
  {\bibfield  {journal} {\bibinfo  {journal} {Nucl. Phys. B}\ }\textbf
  {\bibinfo {volume} {783}},\ \bibinfo {pages} {90 } (\bibinfo {year}
  {2007})}\BibitemShut {NoStop}%
\bibitem [{\citenamefont {Engel}\ and\ \citenamefont
  {Men{\'{e}}ndez}(2017)}]{Engel2017}%
  \BibitemOpen
  \bibfield  {author} {\bibinfo {author} {\bibfnamefont {J.}~\bibnamefont
  {Engel}}\ and\ \bibinfo {author} {\bibfnamefont {J.}~\bibnamefont
  {Men{\'{e}}ndez}},\ }\bibfield  {title} {\bibinfo {title} {Status and future
  of nuclear matrix elements for neutrinoless double-beta decay: a review},\
  }\href {https://doi.org/10.1088/1361-6633/aa5bc5} {\bibfield  {journal}
  {\bibinfo  {journal} {Reports on Progress in Physics}\ }\textbf {\bibinfo
  {volume} {80}},\ \bibinfo {pages} {046301} (\bibinfo {year}
  {2017})}\BibitemShut {NoStop}%
\bibitem [{\citenamefont {Šimkovic}\ and\ \citenamefont
  {Faessler}(2002)}]{Simkovic2002}%
  \BibitemOpen
  \bibfield  {author} {\bibinfo {author} {\bibfnamefont {F.}~\bibnamefont
  {Šimkovic}}\ and\ \bibinfo {author} {\bibfnamefont {A.}~\bibnamefont
  {Faessler}},\ }\bibfield  {title} {\bibinfo {title} {Distinguishing the
  $0\nu\beta\beta$-decay mechanisms},\ }\href
  {https://doi.org/https://doi.org/10.1016/S0146-6410(02)00125-4} {\bibfield
  {journal} {\bibinfo  {journal} {Prog. Part. Nucl. Phys.}\ }\textbf {\bibinfo
  {volume} {48}},\ \bibinfo {pages} {201} (\bibinfo {year} {2002})}\BibitemShut
  {NoStop}%
\bibitem [{\citenamefont {Abgrall}\ \emph
  {et~al.}(2021{\natexlab{a}})\citenamefont {Abgrall} \emph
  {et~al.}}]{Legend:pCDR}%
  \BibitemOpen
  \bibfield  {author} {\bibinfo {author} {\bibfnamefont {N.}~\bibnamefont
  {Abgrall}} \emph {et~al.} (\bibinfo {collaboration} {LEGEND Collaboration}),\
  }\href {https://doi.org/10.48550/arXiv.2107.11462} {\bibinfo {title}
  {{LEGEND}-1000 preconceptual design report}} (\bibinfo {year}
  {2021}{\natexlab{a}}),\ \Eprint {https://arxiv.org/abs/2107.11462}
  {arXiv:2107.11462 [nucl-ex]} \BibitemShut {NoStop}%
\bibitem [{\citenamefont {Abgrall}\ \emph {et~al.}(2025)\citenamefont {Abgrall}
  \emph {et~al.}}]{mjd:instrumentation2025}%
  \BibitemOpen
  \bibfield  {author} {\bibinfo {author} {\bibfnamefont {N.}~\bibnamefont
  {Abgrall}} \emph {et~al.} (\bibinfo {collaboration} {\textsc{Majorana}
  Collaboration}),\ }\bibfield  {title} {\bibinfo {title} {The \textsc{Majorana
  Demonstrator} experiment's construction, commissioning, and performance},\
  }\bibfield  {journal} {\bibinfo  {journal} {Phys. Rev. C}\ }\href
  {https://doi.org/10.48550/arXiv.2501.02060} {10.48550/arXiv.2501.02060}
  (\bibinfo {year} {2025}),\ \bibinfo {note} {(Submitted)}\BibitemShut
  {NoStop}%
\bibitem [{\citenamefont {Agostini}\ \emph {et~al.}(2018)\citenamefont
  {Agostini} \emph {et~al.}}]{gerda2018}%
  \BibitemOpen
  \bibfield  {author} {\bibinfo {author} {\bibfnamefont {M.}~\bibnamefont
  {Agostini}} \emph {et~al.} (\bibinfo {collaboration} {{GERDA}
  Collaboration}),\ }\bibfield  {title} {\bibinfo {title} {Upgrade for phase ii
  of the {GERDA} experiment},\ }\bibfield  {journal} {\bibinfo  {journal} {Eur.
  Phys. J. C}\ }\textbf {\bibinfo {volume} {78}},\ \href
  {https://doi.org/10.1140/epjc/s10052-018-5812-2}
  {10.1140/epjc/s10052-018-5812-2} (\bibinfo {year} {2018})\BibitemShut
  {NoStop}%
\bibitem [{\citenamefont {Arnquist}\ \emph
  {et~al.}(2023{\natexlab{a}})\citenamefont {Arnquist} \emph
  {et~al.}}]{mjd2023}%
  \BibitemOpen
  \bibfield  {author} {\bibinfo {author} {\bibfnamefont {I.~J.}\ \bibnamefont
  {Arnquist}} \emph {et~al.} (\bibinfo {collaboration} {\textsc{Majorana}
  Collaboration}),\ }\bibfield  {title} {\bibinfo {title} {Final result of the
  \textsc{Majorana Demonstrator}'s search for neutrinoless
  double-$\ensuremath{\beta}$ decay in $^{76}\mathrm{Ge}$},\ }\href
  {https://doi.org/10.1103/PhysRevLett.130.062501} {\bibfield  {journal}
  {\bibinfo  {journal} {Phys. Rev. Lett.}\ }\textbf {\bibinfo {volume} {130}},\
  \bibinfo {pages} {062501} (\bibinfo {year} {2023}{\natexlab{a}})}\BibitemShut
  {NoStop}%
\bibitem [{\citenamefont {Agostini}\ \emph {et~al.}(2020)\citenamefont
  {Agostini} \emph {et~al.}}]{gerda2020}%
  \BibitemOpen
  \bibfield  {author} {\bibinfo {author} {\bibfnamefont {M.}~\bibnamefont
  {Agostini}} \emph {et~al.} (\bibinfo {collaboration} {{GERDA}
  Collaboration}),\ }\bibfield  {title} {\bibinfo {title} {Final results of
  {GERDA} on the search for neutrinoless double-$\ensuremath{\beta}$ decay},\
  }\href {https://doi.org/10.1103/PhysRevLett.125.252502} {\bibfield  {journal}
  {\bibinfo  {journal} {Phys. Rev. Lett.}\ }\textbf {\bibinfo {volume} {125}},\
  \bibinfo {pages} {252502} (\bibinfo {year} {2020})}\BibitemShut {NoStop}%
\bibitem [{\citenamefont {Arnquist}\ \emph {et~al.}(2021)\citenamefont
  {Arnquist} \emph {et~al.}}]{mjd:ES2021}%
  \BibitemOpen
  \bibfield  {author} {\bibinfo {author} {\bibfnamefont {I.~J.}\ \bibnamefont
  {Arnquist}} \emph {et~al.} (\bibinfo {collaboration} {\textsc{Majorana}
  Collaboration}),\ }\bibfield  {title} {\bibinfo {title} {Search for
  double-$\ensuremath{\beta}$ decay of $^{76}\mathrm{Ge}$ to excited states of
  $^{76}\mathrm{Se}$ with the \textsc{Majorana Demonstrator}},\ }\href
  {https://doi.org/10.1103/PhysRevC.103.015501} {\bibfield  {journal} {\bibinfo
   {journal} {Phys. Rev. C}\ }\textbf {\bibinfo {volume} {103}},\ \bibinfo
  {pages} {015501} (\bibinfo {year} {2021})}\BibitemShut {NoStop}%
\bibitem [{\citenamefont {Agostini}\ \emph {et~al.}(2015)\citenamefont
  {Agostini} \emph {et~al.}}]{gerda:ES}%
  \BibitemOpen
  \bibfield  {author} {\bibinfo {author} {\bibfnamefont {M.}~\bibnamefont
  {Agostini}} \emph {et~al.} (\bibinfo {collaboration} {{GERDA}
  Collaboration}),\ }\bibfield  {title} {\bibinfo {title}
  {{2$\nu$$\beta$$\beta$} decay of {$^{76}$Ge} into excited states with
  {{GERDA} phase I}},\ }\href {https://doi.org/10.1088/0954-3899/42/11/115201}
  {\bibfield  {journal} {\bibinfo  {journal} {Journal of Physics G: Nuclear and
  Particle Physics}\ }\textbf {\bibinfo {volume} {42}},\ \bibinfo {pages}
  {115201} (\bibinfo {year} {2015})}\BibitemShut {NoStop}%
\bibitem [{\citenamefont {Hoppe}\ \emph {et~al.}(2014)\citenamefont {Hoppe},
  \citenamefont {Aalseth}, \citenamefont {Farmer}, \citenamefont {Hossbach},
  \citenamefont {Liezers}, \citenamefont {Miley}, \citenamefont {Overman},\
  and\ \citenamefont {Reeves}}]{mjd:efcu}%
  \BibitemOpen
  \bibfield  {author} {\bibinfo {author} {\bibfnamefont {E.}~\bibnamefont
  {Hoppe}}, \bibinfo {author} {\bibfnamefont {C.}~\bibnamefont {Aalseth}},
  \bibinfo {author} {\bibfnamefont {O.}~\bibnamefont {Farmer}}, \bibinfo
  {author} {\bibfnamefont {T.}~\bibnamefont {Hossbach}}, \bibinfo {author}
  {\bibfnamefont {M.}~\bibnamefont {Liezers}}, \bibinfo {author} {\bibfnamefont
  {H.}~\bibnamefont {Miley}}, \bibinfo {author} {\bibfnamefont
  {N.}~\bibnamefont {Overman}},\ and\ \bibinfo {author} {\bibfnamefont
  {J.}~\bibnamefont {Reeves}},\ }\bibfield  {title} {\bibinfo {title}
  {Reduction of radioactive backgrounds in electroformed copper for
  ultra-sensitive radiation detectors},\ }\href
  {https://doi.org/https://doi.org/10.1016/j.nima.2014.06.082} {\bibfield
  {journal} {\bibinfo  {journal} {Nucl. Instrum. Methods Phys. Res. A}\
  }\textbf {\bibinfo {volume} {764}},\ \bibinfo {pages} {116} (\bibinfo {year}
  {2014})}\BibitemShut {NoStop}%
\bibitem [{\citenamefont {Abgrall}\ \emph {et~al.}(2016)\citenamefont {Abgrall}
  \emph {et~al.}}]{mjd:assay}%
  \BibitemOpen
  \bibfield  {author} {\bibinfo {author} {\bibfnamefont {N.}~\bibnamefont
  {Abgrall}} \emph {et~al.} (\bibinfo {collaboration} {\textsc{Majorana}
  Collaboration}),\ }\bibfield  {title} {\bibinfo {title} {The \textsc{Majorana
  Demonstrator} radioassay program},\ }\href
  {https://doi.org/https://doi.org/10.1016/j.nima.2016.04.070} {\bibfield
  {journal} {\bibinfo  {journal} {Nucl. Instrum. Methods Phys. Res. A}\
  }\textbf {\bibinfo {volume} {828}},\ \bibinfo {pages} {22} (\bibinfo {year}
  {2016})}\BibitemShut {NoStop}%
\bibitem [{\citenamefont {Bugg}\ \emph {et~al.}(2014)\citenamefont {Bugg},
  \citenamefont {Efremenko},\ and\ \citenamefont {Vasilyev}}]{mjd:muonveto}%
  \BibitemOpen
  \bibfield  {author} {\bibinfo {author} {\bibfnamefont {W.}~\bibnamefont
  {Bugg}}, \bibinfo {author} {\bibfnamefont {Y.}~\bibnamefont {Efremenko}},\
  and\ \bibinfo {author} {\bibfnamefont {S.}~\bibnamefont {Vasilyev}},\
  }\bibfield  {title} {\bibinfo {title} {Large plastic scintillator panels with
  {WLS} fiber readout: Optimization of components},\ }\href
  {https://doi.org/https://doi.org/10.1016/j.nima.2014.05.055} {\bibfield
  {journal} {\bibinfo  {journal} {Nucl. Instrum. Methods Phys. Res. A}\
  }\textbf {\bibinfo {volume} {758}},\ \bibinfo {pages} {91 } (\bibinfo {year}
  {2014})}\BibitemShut {NoStop}%
\bibitem [{\citenamefont {Abgrall}\ \emph
  {et~al.}(2017{\natexlab{a}})\citenamefont {Abgrall} \emph
  {et~al.}}]{mjd:muonflux}%
  \BibitemOpen
  \bibfield  {author} {\bibinfo {author} {\bibfnamefont {N.}~\bibnamefont
  {Abgrall}} \emph {et~al.} (\bibinfo {collaboration} {\textsc{Majorana}
  Collaboration}),\ }\bibfield  {title} {\bibinfo {title} {Muon flux
  measurements at the davis campus of the sanford underground research facility
  with the \textsc{Majorana Demonstrator} veto system},\ }\href
  {https://doi.org/https://doi.org/10.1016/j.astropartphys.2017.01.013}
  {\bibfield  {journal} {\bibinfo  {journal} {Astroparticle Physics}\ }\textbf
  {\bibinfo {volume} {93}},\ \bibinfo {pages} {70} (\bibinfo {year}
  {2017}{\natexlab{a}})}\BibitemShut {NoStop}%
\bibitem [{\citenamefont {Abgrall}\ \emph
  {et~al.}(2017{\natexlab{b}})\citenamefont {Abgrall} \emph
  {et~al.}}]{mjd:calibration}%
  \BibitemOpen
  \bibfield  {author} {\bibinfo {author} {\bibfnamefont {N.}~\bibnamefont
  {Abgrall}} \emph {et~al.} (\bibinfo {collaboration} {\textsc{Majorana}
  Collaboration}),\ }\bibfield  {title} {\bibinfo {title} {The \textsc{Majorana
  Demonstrator} calibration system},\ }\href
  {https://doi.org/https://doi.org/10.1016/j.nima.2017.08.005} {\bibfield
  {journal} {\bibinfo  {journal} {Nucl. Instrum. Methods Phys. Res. A}\
  }\textbf {\bibinfo {volume} {872}},\ \bibinfo {pages} {16} (\bibinfo {year}
  {2017}{\natexlab{b}})}\BibitemShut {NoStop}%
\bibitem [{\citenamefont {Arnquist}\ \emph
  {et~al.}(2023{\natexlab{b}})\citenamefont {Arnquist} \emph
  {et~al.}}]{mjd:calibrationprocedure}%
  \BibitemOpen
  \bibfield  {author} {\bibinfo {author} {\bibfnamefont {I.}~\bibnamefont
  {Arnquist}} \emph {et~al.} (\bibinfo {collaboration} {\textsc{Majorana}
  collaboration}),\ }\bibfield  {title} {\bibinfo {title} {Energy calibration
  of germanium detectors for the \textsc{Majorana Demonstrator}},\ }\href
  {https://doi.org/10.1088/1748-0221/18/09/P09023} {\bibfield  {journal}
  {\bibinfo  {journal} {Journal of Instrumentation}\ }\textbf {\bibinfo
  {volume} {18}}\bibinfo  {number} { (09)},\ \bibinfo {pages}
  {P09023}}\BibitemShut {NoStop}%
\bibitem [{\citenamefont {Arnquist}\ \emph
  {et~al.}(2022{\natexlab{a}})\citenamefont {Arnquist} \emph
  {et~al.}}]{mjd:electronics}%
  \BibitemOpen
\bibfield  {number} {  }\bibfield  {author} {\bibinfo {author} {\bibfnamefont
  {I.~J.}\ \bibnamefont {Arnquist}} \emph {et~al.} (\bibinfo {collaboration}
  {\textsc{Majorana} Collaboration}),\ }\bibfield  {title} {\bibinfo {title}
  {The \textsc{Majorana Demonstrator} readout electronics system},\ }\href
  {https://doi.org/10.1088/1748-0221/17/05/T05003} {\bibfield  {journal}
  {\bibinfo  {journal} {JINST}\ }\textbf {\bibinfo {volume} {17}}\bibinfo
  {number} { (05)},\ \bibinfo {pages} {T05003}}\BibitemShut {NoStop}%
\bibitem [{\citenamefont {Alvis}\ \emph
  {et~al.}(2019{\natexlab{a}})\citenamefont {Alvis} \emph {et~al.}}]{mjd2019}%
  \BibitemOpen
\bibfield  {number} {  }\bibfield  {author} {\bibinfo {author} {\bibfnamefont
  {S.~I.}\ \bibnamefont {Alvis}} \emph {et~al.} (\bibinfo {collaboration}
  {\textsc{Majorana} Collaboration}),\ }\bibfield  {title} {\bibinfo {title}
  {Search for neutrinoless double-$\ensuremath{\beta}$ decay in
  $^{76}\mathrm{Ge}$ with 26 kg yr of exposure from the \textsc{Majorana
  Demonstrator}},\ }\href {https://doi.org/10.1103/PhysRevC.100.025501}
  {\bibfield  {journal} {\bibinfo  {journal} {Phys. Rev. C}\ }\textbf {\bibinfo
  {volume} {100}},\ \bibinfo {pages} {025501} (\bibinfo {year}
  {2019}{\natexlab{a}})}\BibitemShut {NoStop}%
\bibitem [{\citenamefont {Alvis}\ \emph
  {et~al.}(2019{\natexlab{b}})\citenamefont {Alvis} \emph {et~al.}}]{mjd:avse}%
  \BibitemOpen
  \bibfield  {author} {\bibinfo {author} {\bibfnamefont {S.~I.}\ \bibnamefont
  {Alvis}} \emph {et~al.} (\bibinfo {collaboration} {\textsc{Majorana}
  Collaboration}),\ }\bibfield  {title} {\bibinfo {title} {Multisite event
  discrimination for the \textsc{Majorana Demonstrator}},\ }\href
  {https://doi.org/10.1103/PhysRevC.99.065501} {\bibfield  {journal} {\bibinfo
  {journal} {Phys. Rev. C}\ }\textbf {\bibinfo {volume} {99}},\ \bibinfo
  {pages} {065501} (\bibinfo {year} {2019}{\natexlab{b}})}\BibitemShut
  {NoStop}%
\bibitem [{\citenamefont {Arnquist}\ \emph
  {et~al.}(2022{\natexlab{b}})\citenamefont {Arnquist} \emph
  {et~al.}}]{mjd:dcr}%
  \BibitemOpen
  \bibfield  {author} {\bibinfo {author} {\bibfnamefont {I.}~\bibnamefont
  {Arnquist}} \emph {et~al.} (\bibinfo {collaboration} {\textsc{Majorana}
  Collaboration}),\ }\bibfield  {title} {\bibinfo {title} {$\alpha$-event
  characterization and rejection in point-contact {HPGe} detectors},\ }\href
  {https://doi.org/10.1140/epjc/s10052-022-10161-y} {\bibfield  {journal}
  {\bibinfo  {journal} {Eur. Phys. J. C}\ }\textbf {\bibinfo {volume} {82}},\
  \bibinfo {pages} {226} (\bibinfo {year} {2022}{\natexlab{b}})}\BibitemShut
  {NoStop}%
\bibitem [{\citenamefont {Abgrall}\ \emph
  {et~al.}(2021{\natexlab{b}})\citenamefont {Abgrall} \emph
  {et~al.}}]{mjd:adcnonlin}%
  \BibitemOpen
  \bibfield  {author} {\bibinfo {author} {\bibfnamefont {N.}~\bibnamefont
  {Abgrall}} \emph {et~al.} (\bibinfo {collaboration} {\textsc{Majorana}
  Collaboration}),\ }\bibfield  {title} {\bibinfo {title} {{ADC Nonlinearity
  Correction for the \textsc{Majorana Demonstrator}}},\ }\href
  {https://doi.org/10.1109/TNS.2020.3043671} {\bibfield  {journal} {\bibinfo
  {journal} {IEEE Trans. Nucl. Sci.}\ }\textbf {\bibinfo {volume} {68}},\
  \bibinfo {pages} {359} (\bibinfo {year} {2021}{\natexlab{b}})}\BibitemShut
  {NoStop}%
\bibitem [{\citenamefont {Arnquist}\ \emph
  {et~al.}(2023{\natexlab{c}})\citenamefont {Arnquist} \emph
  {et~al.}}]{mjd:chargetrapping}%
  \BibitemOpen
  \bibfield  {author} {\bibinfo {author} {\bibfnamefont {I.~J.}\ \bibnamefont
  {Arnquist}} \emph {et~al.} (\bibinfo {collaboration} {\textsc{Majorana}
  Collaboration}),\ }\bibfield  {title} {\bibinfo {title} {Charge trapping
  correction and energy performance of the \textsc{Majorana Demonstrator}},\
  }\href {https://doi.org/10.1103/PhysRevC.107.045503} {\bibfield  {journal}
  {\bibinfo  {journal} {Phys. Rev. C}\ }\textbf {\bibinfo {volume} {107}},\
  \bibinfo {pages} {045503} (\bibinfo {year} {2023}{\natexlab{c}})}\BibitemShut
  {NoStop}%
\bibitem [{\citenamefont {Heise}(2015)}]{surf}%
  \BibitemOpen
  \bibfield  {author} {\bibinfo {author} {\bibfnamefont {J.}~\bibnamefont
  {Heise}},\ }\bibfield  {title} {\bibinfo {title} {{The Sanford Underground
  Research Facility at Homestake}},\ }\href
  {https://doi.org/10.1088/1742-6596/606/1/012015} {\bibfield  {journal}
  {\bibinfo  {journal} {Journal of Physics: Conference Series}\ }\textbf
  {\bibinfo {volume} {606}},\ \bibinfo {pages} {012015} (\bibinfo {year}
  {2015})}\BibitemShut {NoStop}%
\bibitem [{\citenamefont {Barbeau}\ \emph {et~al.}(2007)\citenamefont
  {Barbeau}, \citenamefont {Collar},\ and\ \citenamefont
  {Tench}}]{Barbeau2007}%
  \BibitemOpen
  \bibfield  {author} {\bibinfo {author} {\bibfnamefont {P.~S.}\ \bibnamefont
  {Barbeau}}, \bibinfo {author} {\bibfnamefont {J.~I.}\ \bibnamefont
  {Collar}},\ and\ \bibinfo {author} {\bibfnamefont {O.}~\bibnamefont
  {Tench}},\ }\bibfield  {title} {\bibinfo {title} {Large-mass ultralow noise
  germanium detectors: performance and applications in neutrino and
  astroparticle physics},\ }\href
  {https://doi.org/10.1088/1475-7516/2007/09/009} {\bibfield  {journal}
  {\bibinfo  {journal} {J. Cosmol. Astropart. Phys.}\ }\textbf {\bibinfo
  {volume} {2007}}\bibinfo  {number} { (09)},\ \bibinfo {pages}
  {009}}\BibitemShut {NoStop}%
\bibitem [{\citenamefont {Cooper}\ \emph {et~al.}(2011)\citenamefont {Cooper},
  \citenamefont {Radford}, \citenamefont {Hausladen},\ and\ \citenamefont
  {Lagergren}}]{Cooper2011}%
  \BibitemOpen
\bibfield  {number} {  }\bibfield  {author} {\bibinfo {author} {\bibfnamefont
  {R.}~\bibnamefont {Cooper}}, \bibinfo {author} {\bibfnamefont
  {D.}~\bibnamefont {Radford}}, \bibinfo {author} {\bibfnamefont
  {P.}~\bibnamefont {Hausladen}},\ and\ \bibinfo {author} {\bibfnamefont
  {K.}~\bibnamefont {Lagergren}},\ }\bibfield  {title} {\bibinfo {title} {A
  novel {HPGe} detector for gamma-ray tracking and imaging},\ }\href
  {https://doi.org/https://doi.org/10.1016/j.nima.2011.10.008} {\bibfield
  {journal} {\bibinfo  {journal} {Nucl. Instrum. Methods Phys. Res. A}\
  }\textbf {\bibinfo {volume} {665}},\ \bibinfo {pages} {25} (\bibinfo {year}
  {2011})}\BibitemShut {NoStop}%
\bibitem [{Can()}]{Canberra:bege}%
  \BibitemOpen
  \href@noop {} {}\bibinfo {note} {Canberra Industries Inc. (now Mirion
  Technologies), 800 Research Parkway Meriden, CT 06450,
  \url{https://www.mirion.com/products/bege-broad-energy-germanium-detectors}}\BibitemShut
  {NoStop}%
\bibitem [{\citenamefont {Arnquist}\ \emph
  {et~al.}(2022{\natexlab{c}})\citenamefont {Arnquist} \emph
  {et~al.}}]{mjd:muonactivity}%
  \BibitemOpen
  \bibfield  {author} {\bibinfo {author} {\bibfnamefont {I.~J.}\ \bibnamefont
  {Arnquist}} \emph {et~al.} (\bibinfo {collaboration} {\textsc{Majorana}
  Collaboration}),\ }\bibfield  {title} {\bibinfo {title} {Signatures of muonic
  activation in the \textsc{Majorana Demonstrator}},\ }\href
  {https://doi.org/10.1103/PhysRevC.105.014617} {\bibfield  {journal} {\bibinfo
   {journal} {Phys. Rev. C}\ }\textbf {\bibinfo {volume} {105}},\ \bibinfo
  {pages} {014617} (\bibinfo {year} {2022}{\natexlab{c}})}\BibitemShut
  {NoStop}%
\bibitem [{\citenamefont {Boswell}\ \emph {et~al.}(2011)\citenamefont {Boswell}
  \emph {et~al.}}]{mjd:mage}%
  \BibitemOpen
  \bibfield  {author} {\bibinfo {author} {\bibfnamefont {M.}~\bibnamefont
  {Boswell}} \emph {et~al.},\ }\bibfield  {title} {\bibinfo {title} {{MaGe-a
  Geant4-Based Monte Carlo Application Framework for Low-Background Germanium
  Experiments}},\ }\href {https://doi.org/10.1109/TNS.2011.2144619} {\bibfield
  {journal} {\bibinfo  {journal} {IEEE Trans. Nucl. Sci.}\ }\textbf {\bibinfo
  {volume} {58}},\ \bibinfo {pages} {1212} (\bibinfo {year}
  {2011})}\BibitemShut {NoStop}%
\bibitem [{\citenamefont {Agostinelli}\ \emph {et~al.}(2003)\citenamefont
  {Agostinelli} \emph {et~al.}}]{geant2003}%
  \BibitemOpen
  \bibfield  {author} {\bibinfo {author} {\bibfnamefont {S.}~\bibnamefont
  {Agostinelli}} \emph {et~al.} (\bibinfo {collaboration} {Geant4
  Collaboration}),\ }\bibfield  {title} {\bibinfo {title} {Geant4—a
  simulation toolkit},\ }\href
  {https://doi.org/https://doi.org/10.1016/S0168-9002(03)01368-8} {\bibfield
  {journal} {\bibinfo  {journal} {Nuclear Instruments and Methods in Physics
  Research Section A: Accelerators, Spectrometers, Detectors and Associated
  Equipment}\ }\textbf {\bibinfo {volume} {506}},\ \bibinfo {pages} {250 }
  (\bibinfo {year} {2003})}\BibitemShut {NoStop}%
\bibitem [{\citenamefont {Ponkratenko}\ \emph {et~al.}(2000)\citenamefont
  {Ponkratenko}, \citenamefont {Tretyak},\ and\ \citenamefont
  {Zdesenko}}]{Ponkratenko2000}%
  \BibitemOpen
  \bibfield  {author} {\bibinfo {author} {\bibfnamefont {O.~A.}\ \bibnamefont
  {Ponkratenko}}, \bibinfo {author} {\bibfnamefont {V.~I.}\ \bibnamefont
  {Tretyak}},\ and\ \bibinfo {author} {\bibfnamefont {Y.~G.}\ \bibnamefont
  {Zdesenko}},\ }\bibfield  {title} {\bibinfo {title} {Event generator {DECAY4}
  for simulating double-beta processes and decays of radioactive nuclei},\
  }\href {https://doi.org/10.1134/1.855784} {\bibfield  {journal} {\bibinfo
  {journal} {Physics of Atomic Nuclei}\ }\textbf {\bibinfo {volume} {63}},\
  \bibinfo {pages} {1282} (\bibinfo {year} {2000})}\BibitemShut {NoStop}%
\bibitem [{\citenamefont {Reine}(2023)}]{reine:2023}%
  \BibitemOpen
  \bibfield  {author} {\bibinfo {author} {\bibfnamefont {A.}~\bibnamefont
  {Reine}},\ }\emph {\bibinfo {title} {A Radiogenic Background Model for the
  \textsc{Majorana Demonstrator}}},\ \href
  {https://doi.org/https://doi.org/10.17615/kbyr-mz98} {Ph.D. thesis},\
  \bibinfo  {school} {University of North Carolina at Chapel Hill} (\bibinfo
  {year} {2023})\BibitemShut {NoStop}%
\bibitem [{\citenamefont {Haufe}(2023)}]{haufe:2023}%
  \BibitemOpen
  \bibfield  {author} {\bibinfo {author} {\bibfnamefont {C.}~\bibnamefont
  {Haufe}},\ }\emph {\bibinfo {title} {A Study of \textsc{Majorana
  Demonstrator} Backgrounds with Bayesian Statistical Modeling}},\ \href
  {https://doi.org/https://doi.org/10.17615/hfck-nx53} {Ph.D. thesis},\
  \bibinfo  {school} {University of North Carolina at Chapel Hill} (\bibinfo
  {year} {2023})\BibitemShut {NoStop}%
\bibitem [{\citenamefont {Dembinski}\ and\ \citenamefont
  {et~al.}(2020)}]{iminuit}%
  \BibitemOpen
  \bibfield  {author} {\bibinfo {author} {\bibfnamefont {H.}~\bibnamefont
  {Dembinski}}\ and\ \bibinfo {author} {\bibfnamefont {P.~O.}\ \bibnamefont
  {et~al.}},\ }\bibfield  {title} {\bibinfo {title} {scikit-hep/iminuit},\
  }\bibfield  {journal} {\bibinfo  {journal} {Zenodo}\ }\href
  {https://doi.org/10.5281/zenodo.3949207} {10.5281/zenodo.3949207} (\bibinfo
  {year} {2020})\BibitemShut {NoStop}%
\bibitem [{\citenamefont {Aunola}\ and\ \citenamefont
  {Suhonen}(1996)}]{Aunola1996}%
  \BibitemOpen
  \bibfield  {author} {\bibinfo {author} {\bibfnamefont {M.}~\bibnamefont
  {Aunola}}\ and\ \bibinfo {author} {\bibfnamefont {J.}~\bibnamefont
  {Suhonen}},\ }\bibfield  {title} {\bibinfo {title} {Systematic study of beta
  and double beta decay to excited final states},\ }\href
  {https://doi.org/https://doi.org/10.1016/0375-9474(96)00087-5} {\bibfield
  {journal} {\bibinfo  {journal} {Nuclear Physics A}\ }\textbf {\bibinfo
  {volume} {602}},\ \bibinfo {pages} {133} (\bibinfo {year}
  {1996})}\BibitemShut {NoStop}%
\bibitem [{\citenamefont {Stoica}\ and\ \citenamefont
  {Mihut}(1996)}]{Stoica1996}%
  \BibitemOpen
  \bibfield  {author} {\bibinfo {author} {\bibfnamefont {S.}~\bibnamefont
  {Stoica}}\ and\ \bibinfo {author} {\bibfnamefont {I.}~\bibnamefont {Mihut}},\
  }\bibfield  {title} {\bibinfo {title} {Nuclear structure calculations of
  two-neutrino double-beta decay transitions to excited final states},\ }\href
  {https://doi.org/https://doi.org/10.1016/0375-9474(96)00122-4} {\bibfield
  {journal} {\bibinfo  {journal} {Nuclear Physics A}\ }\textbf {\bibinfo
  {volume} {602}},\ \bibinfo {pages} {197} (\bibinfo {year}
  {1996})}\BibitemShut {NoStop}%
\bibitem [{\citenamefont {Toivanen}\ and\ \citenamefont
  {Suhonen}(1997)}]{Toivanen1997}%
  \BibitemOpen
  \bibfield  {author} {\bibinfo {author} {\bibfnamefont {J.}~\bibnamefont
  {Toivanen}}\ and\ \bibinfo {author} {\bibfnamefont {J.}~\bibnamefont
  {Suhonen}},\ }\bibfield  {title} {\bibinfo {title} {Study of several
  double-beta-decaying nuclei using the renormalized proton-neutron
  quasiparticle random-phase approximation},\ }\href
  {https://doi.org/10.1103/PhysRevC.55.2314} {\bibfield  {journal} {\bibinfo
  {journal} {Phys. Rev. C}\ }\textbf {\bibinfo {volume} {55}},\ \bibinfo
  {pages} {2314} (\bibinfo {year} {1997})}\BibitemShut {NoStop}%
\bibitem [{\citenamefont {Coello~P\'erez}\ \emph {et~al.}(2018)\citenamefont
  {Coello~P\'erez}, \citenamefont {Men\'endez},\ and\ \citenamefont
  {Schwenk}}]{Perez2018}%
  \BibitemOpen
  \bibfield  {author} {\bibinfo {author} {\bibfnamefont {E.~A.}\ \bibnamefont
  {Coello~P\'erez}}, \bibinfo {author} {\bibfnamefont {J.}~\bibnamefont
  {Men\'endez}},\ and\ \bibinfo {author} {\bibfnamefont {A.}~\bibnamefont
  {Schwenk}},\ }\bibfield  {title} {\bibinfo {title} {Gamow-teller and
  double-$\ensuremath{\beta}$ decays of heavy nuclei within an effective
  theory},\ }\href {https://doi.org/10.1103/PhysRevC.98.045501} {\bibfield
  {journal} {\bibinfo  {journal} {Phys. Rev. C}\ }\textbf {\bibinfo {volume}
  {98}},\ \bibinfo {pages} {045501} (\bibinfo {year} {2018})}\BibitemShut
  {NoStop}%
\bibitem [{\citenamefont {Barea}\ \emph {et~al.}(2015)\citenamefont {Barea},
  \citenamefont {Kotila},\ and\ \citenamefont {Iachello}}]{Barea2015}%
  \BibitemOpen
  \bibfield  {author} {\bibinfo {author} {\bibfnamefont {J.}~\bibnamefont
  {Barea}}, \bibinfo {author} {\bibfnamefont {J.}~\bibnamefont {Kotila}},\ and\
  \bibinfo {author} {\bibfnamefont {F.}~\bibnamefont {Iachello}},\ }\bibfield
  {title} {\bibinfo {title}
  {$0\ensuremath{\nu}\ensuremath{\beta}\ensuremath{\beta}$ and
  $2\ensuremath{\nu}\ensuremath{\beta}\ensuremath{\beta}$ nuclear matrix
  elements in the interacting boson model with isospin restoration},\ }\href
  {https://doi.org/10.1103/PhysRevC.91.034304} {\bibfield  {journal} {\bibinfo
  {journal} {Phys. Rev. C}\ }\textbf {\bibinfo {volume} {91}},\ \bibinfo
  {pages} {034304} (\bibinfo {year} {2015})}\BibitemShut {NoStop}%
\bibitem [{\citenamefont {Unlu}(2014)}]{Unlu2014}%
  \BibitemOpen
  \bibfield  {author} {\bibinfo {author} {\bibfnamefont {S.}~\bibnamefont
  {Unlu}},\ }\bibfield  {title} {\bibinfo {title} {Quasi random phase
  approximation predictions on two-neutrino double beta decay half-lives to the
  first 2+ state},\ }\href {https://doi.org/10.1088/0256-307X/31/4/042101}
  {\bibfield  {journal} {\bibinfo  {journal} {Chinese Physics Letters}\
  }\textbf {\bibinfo {volume} {31}},\ \bibinfo {pages} {042101} (\bibinfo
  {year} {2014})}\BibitemShut {NoStop}%
\bibitem [{\citenamefont {Kostensalo}\ \emph {et~al.}(2022)\citenamefont
  {Kostensalo}, \citenamefont {Suhonen},\ and\ \citenamefont
  {Zuber}}]{Kostensalo2022}%
  \BibitemOpen
  \bibfield  {author} {\bibinfo {author} {\bibfnamefont {J.}~\bibnamefont
  {Kostensalo}}, \bibinfo {author} {\bibfnamefont {J.}~\bibnamefont
  {Suhonen}},\ and\ \bibinfo {author} {\bibfnamefont {K.}~\bibnamefont
  {Zuber}},\ }\bibfield  {title} {\bibinfo {title} {The first large-scale
  shell-model calculation of the two-neutrino double beta decay of {$^{76}$Ge}
  to the excited states in {$^{76}$Se}},\ }\href
  {https://doi.org/https://doi.org/10.1016/j.physletb.2022.137170} {\bibfield
  {journal} {\bibinfo  {journal} {Phys. Lett. B}\ }\textbf {\bibinfo {volume}
  {831}},\ \bibinfo {pages} {137170} (\bibinfo {year} {2022})}\BibitemShut
  {NoStop}%
\end{thebibliography}%
